\documentclass[prd,aps,superscriptaddress,preprintnumbers,eqsecnum,nofootinbib,notitlepage]{revtex4-2}
\usepackage[utf8]{inputenc}
\usepackage{amsfonts,amsmath,amssymb,bm,subfigure}
\usepackage{multirow,hyperref}
\usepackage{xcolor}
\usepackage{graphicx}
\usepackage{comment}
\usepackage{orcidlink}

\begin{document}
\baselineskip=12pt

\title{Theory of interacting vector dark energy and fluid}

\author{Masroor C. Pookkillath\orcidlink{0000-0002-7199-8037}}
\affiliation{Centre for Theoretical Physics and Natural Philosophy, Mahidol University, Nakhonsawan Campus,  Phayuha Khiri, Nakhonsawan 60130, Thailand}

\author{Kazuya Koyama\orcidlink{0000-0001-6727-6915}}
\affiliation{Institute of Cosmology \& Gravitation, University of Portsmouth, Dennis Sciama Building, Burnaby Road, Portsmouth, PO1 3FX, United Kingdom}

\date{\today}
\begin{abstract}
In this work, we study interaction between dark energy and dark matter, where dark energy is described by a massive vector field, and dark matter is modelled as a fluid. We present a new interaction term, which affects only perturbations and can give interesting phenomenology. Then we present a general Lagrangian for the interacting vector dark energy with dark matter. For the dark energy, we choose Proca theory  with $G_{3}$ term to study its phenomenological consequence. For this model, we explore both background and perturbation dynamics. We also present the no-ghost condition for tensor modes, vector modes and scalar modes. Subsequently, we also study the evolution of the overdensities of both baryon and cold dark matter in the high$-k$ limit. We show that the effective gravitational coupling is modified for cold dark matter and baryon. We also choose a simple concrete model and numerically show a suppression of the effective gravitational coupling for cold dark matter. 
However, in this simple model, the suppression of the effective gravitational coupling does not result in a suppression of the matter overdensity compared to that in the $\Lambda$CDM model due to the modified background expansion. 
\end{abstract}
\maketitle

\section{Introduction}
Our present understanding of the universe apprises us that the universe is composed of 68.3\% of Dark Energy (DE), which is responsible for the accelerated expansion today, 26.8\% of Dark Matter (DM), which is an unknown pressure less fluid responsible for dynamics of galaxies and structure formation and 4.9\% of the known matter. We can model our universe assuming General Relativity (GR) as the theory of gravity together with a cosmological constant $\Lambda$ and a pressureless fluid, called cold dark matter (CDM), and baryons~\cite{Planck:2018vyg}. 

Although this model needs only minimal number of (six) parameters to fit the cosmological observation, it does not explain why the value of the cosmological constant is very small, known as the cosmological constant problem~\cite{Weinberg:1988cp}, and why it starts dominating over the other ingredients in very low redshift, known as the cosmic coincidence problem~\cite{Weinberg:1988cp,Martin:2012bt}. To address these crucial questions mentioned above, there have been attempts to explain the phenomenon of DE, with a dynamical origin, where DE is modelled with a scalar field, such as quintessence~\cite{Wetterich:1987fm, Ratra:1987rm} and k-essence~\cite{Armendariz-Picon:2000nqq, Armendariz-Picon:2000ulo}. 

The scalar field can be coupled with other matter fields. Hence, the dynamical DE is extended to the interacting dark sectors. These ideas have been discussed in the literature, for example,~\cite{Wetterich:1994bg, Amendola:1999er, Pettorino:2008ez}. It is interesting to note that this interaction will have an impact on the evolution of the background, as well as on the large-scale structure~\cite{Brax:2009ab}, which can be used to understand the features of the DE. In~\cite{Baldi:2010vv}, a non-linear study of interacting DE is also carried out.

In recent years, there have been notable tensions/discrepancies in the estimation of certain parameters of the universe modelled by $\Lambda$CDM. The most notable one is the Hubble tension $H_{0}$~\cite{DiValentino:2021izs}(see references there in for different solutions). The measured value of $H_{0}$ from the low-redshift observation and the estimated value from the high-redshift observations, such as cosmic microwave background radiation, are of $ \sim 5 \sigma$ tension~\cite{Riess:2021jrx}. The second discrepancy is the mismatch in the estimation of the $S_{8}$ parameter from CMB and large-scale structure observations, which has a tension around $ \sim 3 \sigma$~\cite{DiValentino:2020vvd}. There is also a report of other anomalies, such as the $\Omega_{k}$ anomaly~\cite{DiValentino:2019qzk,2021PhRvD.103d1301H}, the lensing anomaly~\cite{Motloch:2018pjy}. Furthermore, the results from the latest DESI observation, show that, the dark energy could be dynamical~\cite{DESI:2024mwx}. This shows that, assuming there are no systematic errors in the observations, the $\Lambda$CDM description of our universe could be an approximation to the realistic universe. The unsatisfactory explanation of the cosmological constant problem and the coincidence problem together with the growing anomaly motivates us to explore cosmologies beyond the standard $\Lambda$CDM cosmology. 

To address the $H_{0}$ tension, there have been many proposals that have been growing recently. In general, all these proposals fall generally into two categories: 1) high redshift modification, that is, introduction of early DE (EDE) or modified gravity in high redshift~\cite{Poulin:2018cxd, Braglia:2020auw}, 2) low redshift modification by introducing dynamical dark energy or modified gravity~\cite{DeFelice:2020sdq, DeFelice:2020cpt}. Several interacting dark sector scenarios were also considered in both early and late time modifications~\cite{Bernui:2023byc, Karwal:2021vpk}. 

Of these two proposals, EDE is more promising in addressing the Hubble tension. This also introduces a new coincidence problem that we will have to explain exactly why at that redshift (matter-radiation equality) we have a slight domination of the DE. Recently, an interacting early dark energy was introduced in~\cite{CarrilloGonzalez:2020oac, Karwal:2021vpk} to address this new coincidence problem. This particular approach, that is, the interacting EDE can also address the worsening of the $S_{8}$ tension, while introducing the EDE to resolve the Hubble tension~\footnote{It has been noticed in~\cite{Poulin:2023lkg}, this increase in the $S_{8}$, with introducing EDE to solve $H_{0}$ tension is due to the degeneracy in the $\omega_{\rm cdm}$.}. This motivates us to find models of the early interacting dark sector that can address the tensions of both $H_{0}$ and $S_{8}$ tension simultaneously. 

The modelling of dynamical dark energy and modified gravity has developed enough from simple scalar field models like quintessence and k-essence or more complicated theories like Horndeski theories, which adds additional features like modifying the effective Newton constant $G_{\rm eff}$ at the level of the perturbation. On top of Horndeski theories, there have also been studies to include interactions with DM fluids, for example~\cite{Kase:2019veo, Liu:2023mwx}. These theories were also found to have interesting phenomenological features, such as suppressing $G_{\rm eff}$ to reduce the $S_{8}$ tension on top of Hubble tension~\cite{DiValentino:2017iww, DiValentino:2019ffd}. 

The model of dark energy is not limited to scalar fields. In the literature, there has been exploration to check if DE phenomenology is vector like. It is reported in~\cite{Armendariz-Picon:2004say}, that vector DE can have interesting phenomenology like, it can violate the dominant dark energy conditions, $w_{\rm DE} < -1$, without introducing any ghostlike kinetic term. It can also have tracking attractors, similar to the quintessence trackers. A vector like quintessence has been introduced in~\cite{Kiselev:2004py}. In~\cite{Zimdahl:2000zm}, it is suggested that a time like vector can cause for accelerated expansion of the universe. The vector field is also used to explain inflation~\cite{Ford:1989me, Golovnev:2008cf}. 

The massive vector field is later generalized to include derivative self-interactions, known as the generalized Proca theory~\cite{Heisenberg:2014rta,BeltranJimenez:2016rff}. In this theory, the number of propagating degrees of freedom is still three. The construction of this theory is motivated by the generalization of linearized massive gravity theory, the Fierz-Pauli action in~\cite{deRham:2010ik} \footnote{Massive gravity is another modified gravity theory which can explain accelerated expansion of the universe~\cite{deRham:2014zqa}. Viable theories of massive gravity, its extension and its cosmology are explored in~\cite{DeFelice:2015hla, Comelli:2015ksa, DeFelice:2021trp, DeFelice:2022mcd, DeFelice:2023bwq}}. This derivative self-interacting action has been further extended to the Proca-Nuevo theory in~\cite{deRham:2020yet}. 

The cosmology of the generalized Proca theory has been explored in~\cite{DeFelice:2016yws}, and that of Proca-Nuevo and its extension in~\cite{deRham:2021efp}. For generalized Proca theory, it is shown that the effective gravitational coupling for the perturbation can be modified, and it is confronted with the data and found constraints on the parameters~\cite{deFelice:2017paw}. Later, a full confrontation with the CMB data and other large-scale structure data is carried out, and it is shown that it can reduce the Hubble tension~\cite{DeFelice:2020sdq}. It has also been noticed in~\cite{DeFelice:2020sdq}, when adding large-scale structural data, the Hubble tension is not completely removed. 

The vector dark energy that interacts with cold dark matter is introduced in~\cite{Nakamura:2019phn} (also~\cite{Gomez:2020sfz, Gomez:2022okq}). In~\cite{Nakamura:2019phn}, the possibility of an interaction of the form $L_{\rm int} = -Q f(X)\rho_{c}$, where, $X = - 1/2A_{\mu}A^{\mu}$, and $A_{\mu}$ is the massive vector field was explored. This is similar to the interaction that has been studied in the context of scalar dark energy $L_{\rm int} = Q\dot{\phi}\rho_{c}$ in~\cite{Wetterich:1994bg, Amendola:1999er}. In~\cite{Nakamura:2019phn}, they have included the generalized cubic-order Proca term $G_{3}(X)$ and shown that it can allow the phantom dark energy equation of state. These models have modified the effective gravitational constant at the linear perturbation due to the presence of the $G_{3}(X)$ term. 

In a vector dark energy model that interacts with momentum transfer, is developed in~\cite{DeFelice:2020icf}, where they have considered the interaction term $\mathcal{Z} = -u_{\alpha}A^{\alpha}$. Where, $u_{\alpha}$ is four velocity of the cold dark matter. They derived a general expression for the effective gravitational coupling for the baryon and cold dark matter in the high$-k$ limit and found that there could be a suppression of growth of matter in the universe. This study is similar to its scalar counterpart, which is studied in~\cite{Pourtsidou:2013nha}, with an interaction term $L_{\rm int} = \nabla_{\alpha}\phi u^{\alpha}$. 

In this present work, we ask the question, when we consider the dark energy to be driven by a massive vector field, what the possible interaction terms with cold dark matter are? Indeed, there exist interaction terms other than $\mathcal{Z} = -u_{\alpha}A^{\alpha}$, that have been considered in~\cite{DeFelice:2020icf}. It is also noticed that, we can have parity violating interactions between the vector dark energy and cold dark matter. We then extend our study to a particular parity invariant interacting term $\mathcal{E}= - A^{\alpha}F_{\alpha \beta} u^{\beta}$, where the vector dark energy is described by a generalised Proca action with $G_{3}$ term. We also show that in the high $k$ limit, this particular interaction term can modify the effective gravitational coupling $G_{\rm eff}$ for cold dark matter. We also noted that to have this effect, we do need the $G_{3}(X)$ term for vector dark energy action. This was also the case for the interaction $\mathcal{Z}$, where we need the $G_{3}$ term to have the distinctive features of interaction in the high $k$ limit. This model is interesting in the context of the latest result from DESI collaboration, which provided hints for an expansion history different from that of $\Lambda$CDM cosmology~\cite{DESI:2024mwx}. 

For the cold dark matter action, we will follow a similar approach taken in~\cite{Pourtsidou:2013nha} in the context of interacting scalar dark energy. The cold dark matter fluid is described by a ``pull-back" formalism~\cite{Andersson:2006nr}. However, we will show that even in the context of the scalar dark energy in general relativity, the tensor mode acquires a mass term, which is physically not correct, if a conservation term in the action is not introduced. 

This paper is organized as follows. Section~\ref{sec:int_fluid_des} we derive the possible interaction terms including parity violating interaction terms. Then in section~\ref{sec:interating_lagrangian}, we introduce a general interacting Lagrangian, which is not parity violating, and the matter (including cold dark matter) Lagrangian. In Section~\ref{sec:cosmology}, we proceed to study cosmology, to derive background equations of motion, and find the ghost conditions for the tensor, vector and scalar modes. To study cosmology, we have chosen a generalised Proca action (up to $G_{3}$ term) to describe dark energy. Then, in Section~\ref{sec:linear_perturbation_eom}, we find the linear equations of motion. Subsequently, we derive the evolution of baryonic and cold dark matter overdensities in the high $k$ limit in Section~\ref{sec:QSA}, and we show that the cold dark matter overdensity, $\delta_{c}$ equation and baryonic overdenisty, $\delta_{b}$ equation are modified because of the interaction. In Section~\ref{sec:con_model}, we choose a simple concrete model, and we show that there is a suppression of effective gravitational coupling of cold dark matter. We also numerically solve the matter overdensity and show that this simple concrete model does not suppress the matter overdensity in comparison with $\Lambda$CDM. We then present our conclusion in~\ref{sec:conclu}. In Appendix~\ref{appendix:fluid_action}, we present a fluid description that we used to model cold dark matter. 

\section{Interacting fluid description}\label{sec:int_fluid_des}

The theory of an interacting fluid with the scalar field (dark energy) is discussed in~\cite{Pourtsidou:2013nha}, where the cold dark matter fluid is described by a ``pull-back" formalism (for the details of pull-back formalism, refer to~\cite{Andersson:2006nr}). For the interacting fluid with a scalar field dark energy, the general action is
\begin{equation}
    L = L(n_{\alpha \beta \gamma}, u^{\mu}, \phi, \nabla_{\nu} \phi)\,,
    \label{eq:scalar_int_lag}
\end{equation}
where $n_{\alpha \beta \gamma}$ is a full anti-symmetric tensor, which is dual to the number density of the dark matter fluid $n^{\alpha}$,
\begin{equation}\label{eq:dual_number_density}
    n_{\alpha \beta \gamma} = \epsilon_{\alpha \beta \gamma \delta} n^{\delta}\,,
\end{equation}
and,  $u^{\mu}$ is the fluid velocity and $\phi$ is the scalar field, which describes the dark energy. In this section, we develop an interaction of the fluid with the dark energy field, which originated from a broken $U(1)$ gauge symmetry vector field.

\subsection{Building blocks for interacting vector dark energy}
Since we are interested in a dark energy model as a vector theory, the above Lagrangian (\ref{eq:scalar_int_lag}) can be rewritten assuming no additional changes in the fluid sector.
\begin{equation}
     L = L(n_{\alpha}, u^{\mu}, A_{\nu}, \nabla_{\mu}A_{\nu}, \epsilon_{\alpha \beta \gamma \delta})\,,
\end{equation}
where we have included the $\epsilon_{\alpha \beta \gamma \delta}$ since we will use $F_{\mu \nu}$, which is an anti-symmetric term,
\begin{equation}
    F_{\mu \nu} \equiv \nabla_{\mu} A_{\nu} - \nabla_{\nu}A_{\mu} \,.
\end{equation}
From the combination of $\epsilon_{\alpha \beta \gamma \delta}$ and $F_{\mu \nu}$, we can define the following dual tensor
\begin{equation}
    \tilde{F}_{\alpha \beta} \equiv \epsilon_{\alpha \beta \mu \nu} F^{\mu \nu}\,.
\end{equation}
We can also have the dual fluid number density given by Eq.~(\ref{eq:dual_number_density}). Here, we explicitly present all the possible interactions that could be possible with the vector dark energy with dark matter. 

The idea is to construct invariants from the basic fields that the Lagrangian is depending on. First, let us look at the following invariant from the combinations of $A_{\mu}$ and $u^{\mu}$. 
\begin{equation}\label{eq:calZ}
    \mathcal{Z} \equiv - A_{\mu}u^{\mu}\,.
\end{equation}
This is equivalent to the term considered in~\cite{Pourtsidou:2013nha} with a scalar field.
Since we have the tensor $F_{\mu \nu}$ which is anti-symmetric, we can further have other invariants from the combination of $F_{\mu \nu}$, $n_{\alpha \beta \gamma}$, and $u^{\alpha}$
\begin{equation}
     \mathcal{A}_{p} \equiv F^{\beta \gamma } n_{\alpha \beta \gamma } u^{\alpha }\,.
\end{equation}
The above interaction term is parity violating. Including the $\epsilon$ tensor, we can have the following invariants as well
\begin{eqnarray}
   \mathcal{A}_{1} &\equiv & \epsilon_{\beta \gamma \delta \varepsilon } F^{\beta \gamma } n_{\alpha }{}^{\delta \varepsilon } u^{\alpha } = 4 F_{\alpha \beta} n^{\alpha} u^{\beta}  , \\
   \mathcal{A}_{2} & \equiv & \epsilon_{\alpha \gamma \delta \varepsilon } F^{\beta \gamma } n_{\beta }{}^{\delta \varepsilon } u^{\alpha } = 4 F_{\alpha \beta} n^{\alpha} u^{\beta}\,, \\
   \mathcal{A}_{3} & \equiv & \epsilon_{\beta \gamma \delta \varepsilon } F_{\alpha }{}^{\beta } n^{\gamma \delta \varepsilon } u^{\alpha }= -6 F_{\alpha \beta} n^{\alpha} u^{\beta}\, .
\end{eqnarray}
Instead of $F_{\alpha \beta}$ we can consider its dual $\tilde{F}_{\alpha \beta}$, then we have 
\begin{eqnarray}
    \mathcal{B} & \equiv & n_{\alpha \beta \gamma } \tilde{F}^{\beta \gamma } u^{\alpha}= 4 F_{\alpha \beta} n^{\alpha} u^{\beta}\,.
\end{eqnarray}
Now adding the $\epsilon$ tensor we get 
\begin{eqnarray}
     \mathcal{B}_{p1} & \equiv &\epsilon_{\beta \gamma \delta \varepsilon } n^{\gamma \delta \varepsilon } \tilde{F}_{\alpha }{}^{\beta } u^{\alpha } =  \epsilon_{\beta \gamma \delta \alpha } F^{\gamma \delta } n^{\alpha } u^{\beta } \,, \\
\mathcal{B}_{p2} & \equiv & \epsilon_{\beta \gamma \delta \varepsilon } n_{\alpha }{}^{\delta \varepsilon } \tilde{F}^{\beta \gamma } u^{\alpha } = 4 \epsilon_{\gamma \delta \alpha \beta } F^{\gamma \delta } n^{\alpha } u^{\beta } \,, \\
\mathcal{B}_{p3} & \equiv & \epsilon_{\alpha \gamma \delta \varepsilon } n_{\beta }{}^{\delta \varepsilon } \tilde{F}^{\beta \gamma } u^{\alpha } = 2 \epsilon_{\gamma \delta \alpha \beta } F^{\gamma \delta } n^{\alpha } u^{\beta }\,,
\end{eqnarray}
which are parity violating interaction terms.

Taking into account the terms $F_{\mu \nu}$, $A_{\alpha}$ and $u^{\beta}$ we can build another invariant,
\begin{equation}\label{eq:calE}
    \mathcal{E} \equiv - A^{\alpha } F_{\alpha \beta } u^{\beta }\,.
\end{equation}
The above term is not parity violating. Including the anti-symmetric $\epsilon$ tensor we can have the following invariants
\begin{equation}
    \mathcal{E}_{p} \equiv A^{\alpha } \epsilon_{\alpha \beta \gamma \delta } F^{\gamma \delta } u^{\beta } \equiv A^{\alpha }  \tilde{F}_{\alpha \beta } u^{\beta } \,,
\end{equation}
which is a parity violating interaction term.

Furthermore, we can also have invariant from the combinations from $F_{\alpha \beta}$, $A_{\mu}$, and $n_{\rho \sigma \delta}$
\begin{equation}
    \mathcal{G}_{p} = A^{\alpha } F^{\beta \gamma } n_{\alpha \beta \gamma } \,.
\end{equation}
Now including $\epsilon_{\beta \gamma \delta \varepsilon }$ we have
\begin{eqnarray}
    \mathcal{G}_{1} & \equiv & A^{\alpha } \epsilon_{\beta \gamma \delta \varepsilon } F^{\beta \gamma } n_{\alpha }{}^{\delta \varepsilon }   = -4 A^{\alpha}F_{\alpha \beta} n^{\beta} \,\,, \\ 
    \mathcal{G}_{2} & \equiv &A^{\alpha } \epsilon_{\alpha \gamma \delta \varepsilon } F^{\beta \gamma } n_{\beta }{}^{\delta \varepsilon }  = -2 A^{\alpha} F_{\alpha \beta} n^{\beta} \,,  \\
    \mathcal{G}_{3} & \equiv & A^{\alpha } \epsilon_{\beta \gamma \delta \varepsilon } F_{\alpha }{}^{\beta } n^{\gamma \delta \varepsilon } = 6 A^{\alpha} F_{\alpha \beta} n^{\beta}\,.
\end{eqnarray}
Considering the dual tensor $\tilde{F}_{\alpha \beta}$, then we get
\begin{equation}
     \mathcal{F} \equiv A^{\alpha } n_{\alpha \beta \gamma } \tilde{F}^{\beta \gamma } = -4 A^{\alpha } F_{\alpha \beta } n^{\beta } \,,
\end{equation}
and including $\epsilon$ we get
\begin{eqnarray}
    \mathcal{F}_{p1} & \equiv &  A^{\alpha} \epsilon_{\beta \gamma \delta \varepsilon } n^{\gamma \delta \varepsilon } \tilde{F}_{\alpha }{}^{\beta } = 6 A^{\alpha } \epsilon_{\alpha \gamma \delta \beta } F^{\gamma \delta } n^{\beta } \, ,\\
\mathcal{F}_{p2} & \equiv & A^{\alpha } \epsilon_{\beta \gamma \delta \varepsilon } n_{\alpha}{}^{\delta \varepsilon } \tilde{F}^{\beta \gamma } = -4 A^{\alpha } \epsilon_{\gamma \delta \alpha \beta } F^{\gamma \delta } n^{\beta } \,, \\
\mathcal{F}_{p3} & \equiv & A^{\alpha} \epsilon_{\alpha \gamma \delta \varepsilon } n_{\beta }{}^{\delta \varepsilon } \tilde{F}^{\beta \gamma } = -2 A^{\alpha } \epsilon_{\gamma \delta \alpha \beta } F^{\gamma \delta } n^{\beta } \,.
\end{eqnarray}

The terms with $\epsilon$ are parity violating, so we will not consider them in the following. Then the possible interactions are 
\begin{equation}
    \mathcal{Z} \equiv - A_{\mu}u^{\mu}\,, \qquad \mathcal{A} \equiv F_{\beta \gamma } n^{\beta} u^{\gamma }\,, \qquad  \text{and} \qquad
    \mathcal{E} \equiv - A^{\alpha } F_{\alpha \beta } u^{\beta }\,.
\end{equation}
As for the interaction term $\mathcal{A}$ we can rewrite is as 
\begin{equation}
    \mathcal{A} = F_{\beta \gamma} n^{\beta} u^{\gamma} = n F_{\beta \gamma} u^{\beta} u^{\gamma}\, \qquad \text{where} \qquad n^{\alpha} = n u^{\alpha}\,,
\end{equation}
hence it is zero since $F_{\alpha \beta}$ is antisymmetric. Then the contributing interactions are $\mathcal{Z}$ and $\mathcal{E}$. The first interaction term $\mathcal{Z}$ has already been studied in~\cite{DeFelice:2020icf}, and has shown that, one can suppress cosmic growth. 
 
Apart from this interaction, we have the standard non-interaction terms as 
\begin{equation}
    \mathcal{X} \equiv F_{\mu \nu} F^{\mu \nu}\,, \qquad 
    \mathcal{Y} \equiv  - \frac{1}{2} A_{\mu}A^{\mu}\,,
\end{equation}
where $\mathcal{X}$ is the standard kinetic term and $\mathcal{Y}$ is the mass term in the Proca theory~\cite{Heisenberg:2014rta}. So far, we only consider interactions that are expressed in terms of $F_{\mu \nu}$. We can also consider an interaction term  $A^{\alpha}\nabla_{\alpha}A_{\beta}u^{\beta}$. However, this will give a higher derivative term to the longitudinal mode defined as, $A_{\mu} = \partial_{\mu} \psi$ while the higher derivative term is absent if we construct the interaction in terms of $F_{\mu \nu}$. Hence, we do not consider these kinds of interaction terms in the Lagrangian.

\section{An interacting Lagrangian}\label{sec:interating_lagrangian}
Now we can write the interacting Lagrangian that does not violate the parity as 
\begin{equation}
    L = L\left(\tilde{n}, \mathcal{X}, \mathcal{Y}, \mathcal{Z}, \mathcal{E}\right)\,,
\end{equation}
where
\begin{equation}
    \tilde{n} \equiv n^2 = \frac{1}{3!}n^{\alpha \beta \gamma}n_{\alpha \beta \gamma}\,,
\end{equation}
and is called the fluid number density. 

The Lagrangian that we have considered is completely general. In the following, we assume that the dark energy is described by the $G_{3}$ Proca theory, together with fluid interaction as described above, and explore the physical consequences of the new interaction terms. With matter minimal coupling to gravity, we have the following action.

\begin{equation}
\label{eq:action_example}
    S_{\rm tot} = \int d^{4}x \sqrt{-g} \left[ \frac{M_{\rm P}^2}{2} R - \frac{1}{4}F_{\alpha \beta}F^{\alpha \beta}  + G_{3}(\mathcal{Y})\nabla_{\mu}A^{\mu} +  f(\mathcal{Y}, \mathcal{Z}, \mathcal{E}) \right] +S_{\rm c} + S_{\rm m}\,,
\end{equation}
In the above action Eq.~(\ref{eq:action_example}), the kinetic terms are separated for each propagating mode, and then an interacting action is also added.
Note that we can generalise the action Eq.~(\ref{eq:action_example}) to include the terms beyond $G_{3}$ term in generalized Proca theory~\cite{Heisenberg:2014rta, DeFelice:2016yws}. 

The matter action, we follow the Schutz-Sorkin action~\cite{Schutz:1977df,Brown:1992kc, DeFelice:2009bx, Pookkillath:2019nkn}
\begin{equation}
    S_{\text{m}}	=-\int\text{d}^{4}x\sqrt{-g}\left[\rho_{\text{m}}\left(n_{\text{m}}\right)+J_{\text{m}}^{\mu}\left(\nabla_{\mu}l_{\text{m}}  + \mathcal{A}_{i}{}^{\rm m} \nabla_{\mu} \mathcal{B}^{i}{}^{\rm m}\right)\right]\,,
\end{equation}
where $n_{\text{m}}$ are defined as 
\begin{eqnarray}
    n_{\text{m}} & = & \sqrt{-g_{\mu\nu}J_{\text{m}}^{\mu}J_{\text{m}}^{\nu}}\,,
\end{eqnarray}
and 
\begin{equation}
    J^{\mu} \equiv n_{\rm m} u_{\rm m}^{\mu}\,, \qquad \text{and} \qquad u_{{\rm m} \alpha} u_{\rm m}^{\alpha} = -1 \,.
\end{equation} 

To model the interacting dark sector, we assume that the dark matter is described via the dust fluid as introduced in~\cite{Andersson:2006nr}, which is also used in~\cite{PhysRevD.41.1875, Brown:1992kc, Comer:1993zfa, G_L_Comer_1993, PhysRev.94.1468, Carter1973ElasticPT, Langlois:1997bz}
\begin{equation}
    S_{\rm c} = -\int {\rm d}^4 x \sqrt{-g} \left[ \rho_{c} (\tilde{n}) + n_{c}^{\mu } (\nabla_{\mu }K + \mathcal{A}_{i}{}^{\rm c} \nabla_{\mu} \mathcal{B}^{i}{}^{\rm c})\right]\,,
\end{equation}
The $\rho_{\rm c}(\tilde{n})$ is the energy density of the dark matter component that depends on the square of the number density. This action is different from the action introduced in~\cite{Andersson:2006nr}. We have an additional term,
\begin{equation}
    n_{\rm c}^{\mu}\nabla_{\mu} K\,,
\end{equation}
in other words, we are imposing the $\nabla_{\nu}n_{\rm c}^{\nu} = 0$ with the Lagrangian multiplier $K$. We found that without this term, the perturbation gets an inconsistent tensor-mode propagation. In addition, this term will restrict the interaction to only momentum transfer, since this demands the conservation of the continuity equation. We show in appendix.\ref{appendix:fluid_action} the above action gives the general perfect fluid equations of motion for the linear perturbation.

The last term in the fluid action was introduced in~\cite{Brown:1992kc, Schutz:1977df}. This term is particularly useful to analyse the behaviour of the vector modes, for example in the case of GR one can show that the vector modes decays as $V_{i} \propto 1/a^{2}$ as expected. The field variables $\mathcal{B}_{i}{}^{\rm c, m}$ can be interpreted as Lagrangian coordinates for the perfect fluid. The Lagrangian multiplier $\mathcal{A}_{i}{}^{\rm c, m}$ sets $n^{\mu}\nabla_{\mu} \mathcal{B}^{i}{}^{\rm c} = 0$ and $J^{\mu}\nabla_{\mu} \mathcal{B}^{i}{}^{\rm m} = 0$. 

\section{Cosmology}\label{sec:cosmology}
Now we can study the cosmology of the interacting Lagrangian~(\ref{eq:action_example}) introduced above. In this section, we derive the background and ghost conditions for the tensor, vector, and scalar modes.    

\subsection{Background}
Let us look at the background cosmology. We assume an isotropic and homogeneous flat background, Friedmann-Lema\^{i}tre-Robertson-Walker(FLRW) space-time with linear perturbations,
\begin{eqnarray}
    {\rm d}s^2 = g_{\mu \nu}{\rm d}x^{\mu}{\rm d}x^{\nu} = a^2\left[-(1+2\alpha){\rm d}t^2 + 2\frac{1}{a} \partial_{i}\chi {\rm d}x^{i} {\rm d}t + \left\{(1+2\zeta)\delta_{ij}+ 2\partial_{i}\partial_{j} \left( E/a^2 \right) \right\} {\rm d}x^{i}{\rm d}x^{j} \right]\,.
\end{eqnarray}
For the Proca field $A_{\mu}$ we define
\begin{equation}
    A^{\mu} = \left[ \frac{1}{a}(\phi +\delta \phi) , \frac{1}{a^2}\delta^{ij}\partial_{j}\delta \phi_{s} \right]\,.
\end{equation}
Now for the cold dark matter fluid part we have the fields $n^{\alpha}$ and $K$, which can be defined as
\begin{equation}\label{nu_def}
    n_{c}^{\alpha} = \left[\frac{1}{a}n_{0}(1+\delta n_{0}), \frac{1}{a^2}\delta^{ij}\partial_{j}n_{s}  \right]\,, \qquad K = K_{0} + \delta K\,.
\end{equation}
Finally, we have other matter action for baryons and photons
\begin{eqnarray}
      J_{\text{m}}^{\nu} = \left[\frac{J_{\text{m}}}{a}\left(1+\delta J_{\text{m}}\right),\frac{1}{a^{2}}\delta^{ij}\partial_{j}\delta j_{\text{m}}\right]\,, \qquad
     l_{\text{m}} = \bar{l}_{\text{m}}+\delta l_{\text{m}}\,.   
\end{eqnarray}

For the cold dark matter sector, we have  
\begin{equation}
    n_{0} = \frac{N_{\rm c}}{a^3}\,,
\end{equation}
and
\begin{equation}
    K_{0} = -2 \int a(\eta) n_{0}(\eta) \frac{\partial \rho_{c}}{\partial \tilde{n}} {\rm d} \eta \,. 
\end{equation}
where $N_{\rm c}$ is constant, which is the total number of dark matter particles.

For the matter sector (from $S_{\rm m}$) 
\begin{eqnarray}
    J_{\rm m} &=& \frac{N_{\rm m}}{a^3} \,, \\
    \bar{l}_{\text{m}} &=& -\int_{0}^{t} \rho \left(n_{\rm m} \right)_{,n_{\text{m}}}a \left( \eta \right)\, \text{d}\eta\,.
\end{eqnarray}

Having substituted the matter equations in the action, we can get the Friedmann equations. The first Friedmann  equation, 
\begin{eqnarray}
\label{eq:1stFE}
    3M_{\rm P}^2 H^2 &=& \rho_{\rm m} + \rho_{\rm c} - f  - 3 H \phi^3 G_{3,\mathcal{Y}} - \phi f_{,\mathcal{Z}} - 2 \mathcal{Y}f_{,\mathcal{Y}}\,, 
\end{eqnarray}
where $H = \dot{a}/a^2$. The second Friedmann  equation, 
\begin{eqnarray}
\label{eq:2ndFE}
    2M_{\rm P}^{2}\frac{\dot{H}}{a} & = & - n_{\rm m} \rho_{,n_{\rm m}} + 2 n_{0}^2 \rho_{c,\tilde{n}} + \phi^2 \frac{\dot{\phi}}{a} G_{3,\mathcal{Y}} -  \left( 3 H \phi^3 G_{3,\mathcal{Y}} + \phi f_{,\mathcal{Z}}  + 2 \mathcal{Y}f_{,\mathcal{Y}} \right), \nonumber \\
    &=& -(\rho_{\rm m} + P_{\rm m}) - \rho_{\rm c} 
    -(\rho_{\rm DE} + P_{\rm DE}),\,
\end{eqnarray}
where we have used the relation
\begin{equation}
    n_{\rm m} \rho_{,n_{\rm m}} = \rho_{\rm m} + P_{\rm m}\,, \qquad  2 n_{0}^2 \rho_{c,\tilde{n}} = \rho_{c},
\end{equation}
where $P_{\rm m}$ is the pressure of matter. The density of DE is 
\begin{equation}
    \rho_{\rm DE} = - f  - 3 H \phi^3 G_{3,\mathcal{Y}} - \phi f_{,\mathcal{Z}} - 2 \mathcal{Y}f_{,\mathcal{Y}}\,,
\end{equation}
and the sum of DE density and pressure is 
\begin{equation}
\rho_{\rm DE} + P_{\rm DE}
=- \phi^2 \frac{\dot{\phi}}{a} G_{3,\mathcal{Y}} +  \left( 3 H \phi^3 G_{3,\mathcal{Y}} + \phi f_{,\mathcal{Z}}  + 2 \mathcal{Y}f_{,\mathcal{Y}} \right). 
\end{equation}
Finally, we have the equation for the field $\phi$, which is given as 
\begin{equation}
\label{eq:EQphi}
  3 H \phi^3 G_{3,\mathcal{Y}} + \phi f_{,\mathcal{Z}}  + 2 \mathcal{Y}f_{,\mathcal{Y}} = 0 \,.
\end{equation}

Notice that from the equations of motion for $ \phi$, for the dark energy content $\rho_{\rm DE} + P_{\rm DE} = -\phi^2G_{3,\mathcal{Y}}\tfrac{\dot{\phi}}{a}$, and $\rho_{\rm DE} = -f$. In other words, the dark energy equation of state is given by 
\begin{equation}
    w_{\rm DE} = -1 + \frac{\phi^2 G_{3,\mathcal{Y}}}{f}\frac{\dot{\phi}}{a}\,.
\end{equation}
On top of it, it is interesting to notice that at the background level we do not see the consequences of the interaction, $\mathcal{E}$. However, we need to see how the interactions influence the linear perturbation sector.

\subsection{Perturbation}
In this section we study the ghost instability condition and the Laplace instability conditions for tensor, vector and scalar modes for the theory we are considering Eq.~(\ref{eq:action_example}).

\subsubsection{Tensor modes}
In this subsection, we study the propagation of the tensor mode in this theory. 
We consider the tensor mode perturbation around the FLRW background,
\begin{equation}
\label{eq:metric_tensor}
    {\rm d} s^2 = a(t)^2 \left[-{\rm d}t^2 + \left(\delta_{ij} + \sum_{\lambda = +, \times} \epsilon_{ij}^{\lambda} h_{\lambda} \right) {\rm d}x^{i}{\rm d}x^{j} \right] \,,
\end{equation}
where the tensor mode perturbations  obey standard trace-less and divergence free gauge condition $\delta_{ij}\epsilon_{ij} = 0 $, and $\epsilon_{ij}^{\lambda} \delta^{jl} \partial_{l}h_{\lambda} = 0$, and also it obeys the normalisation condition $\epsilon_{ij}^{+} \epsilon_{kl}^{+} \delta^{ik} \delta^{jl} = \epsilon_{ij}^{\times} \epsilon_{kl}^{\times} \delta^{ik} \delta^{jl} = 1$, with $\epsilon_{ij}^{+}\epsilon_{kl}^{\times}\delta^{ik} \delta^{jl} = 0 $. Now expanding the action Eq.~(\ref{eq:action_example}) with the above metric Eq.~(\ref{eq:metric_tensor}), and substituting the background equations of motion, we get the following reduced action. 
\begin{equation}
    S_{\rm T} = \frac{M_{\rm P}^2}{4} \sum_{\lambda = +, \times} \int {\rm d}^4 x \, a^4 \left[ \frac{\dot{h}_{\lambda}^{2}}{a^2} - \frac{k^2}{a^2} h_{\lambda}^{2} \right] \,,
\end{equation}
where we have changed the action into Fourier space, and $k$ is the Fourier mode. 

Hence, the above expression proves that the tensor mode propagation is not affected.
We note that the new term in the dark matter fluid Lagrangian, $n^{\mu}\nabla_{\mu}K$, which corresponds to the conservation equation, is crucial to obtain this result. Without this term, we can show that the tensor mode acquires a mass term incorrectly.  

\subsubsection{Vector mode}
In this part, we study the no-ghost condition and Laplace instability condition for the vector mode. The theory we are considering is an interaction between dark energy and dark matter, where dark energy described by a massive vector field. We need to check any new non-trivial stability conditions emerges from the propagation of the vector modes. To study the vector mode, we consider a flat-FLRW metric with vector perturbation
\begin{equation}
    {\rm d}s^2 = -a^2 {\rm d} t^{2} + a^2 \delta_{ij} {\rm d}x^{i} {\rm d}x^{j} + 2 a^2 V_{i} {\rm d}x^{i} {\rm d}t \,.
\end{equation}
The vector field $A^{\mu}$ is given by
\begin{equation}
    A^{\mu} = \left( \frac{\phi}{a}, \frac{\delta^{ij}}{a^2} E_{j} \right) \,.
\end{equation}
For the dark matter fluid we define
\begin{equation}
    n_{\rm c}^{\mu} = \left( \frac{n_{0}}{a}, \frac{\delta_{ij}}{a^2}W_{j}{}^{\rm c} \right) \,,
\end{equation}
and for the standard matter we define
\begin{equation}
    J^{\mu}  = \left( \frac{J_{\rm m}}{a}, \frac{\delta_{ij}}{a^2} W_{j}{}^{\rm m} \right) \,.
\end{equation}
The quantities $V_{i}$, $E_{i}$, $W_{i}{}^{\rm c}$, and $W_{i}{}^{\rm m}$ obeys the transverse conditions 
\begin{equation}
    \partial_{i}V^{i} = \partial_{i}E^{i} = \partial_{i}W^{i}{}^{\rm c} = \partial_{i}W^{i}{}^{\rm m} = 0 \,.
\end{equation}
For the convenience of the calculations, we assume the vector fields depends on $t$ and $z$ only, such that we can keep the third component of the vector fields to zero. We also choose, 
\begin{equation}
    \mathcal{A}_{1}{}^{\rm c, m} = \delta \mathcal{A}_{1}{}^{\rm c, m} (t,z)\,, \quad \, \mathcal{A}_{2}{}^{\rm c, m} = \delta \mathcal{A}_{2}{}^{\rm c, m} (t,z)\,, \quad \, \mathcal{B}_{1}{}^{\rm c, m} = x + \delta \mathcal{B}_{1}{}^{\rm c, m} (t,z)\,, \quad \,
    \mathcal{B}_{2}{}^{\rm c, m} = y + \delta \mathcal{B}_{2}{}^{\rm c, m} (t,z)\,.
\end{equation}
Due to the transverse conditions, the third component of $\mathcal{A}_{i}{}^{\rm c,m} \nabla \mathcal{B}^{i}{}^{\rm c, m}$, $\mathcal{A}_{3}\nabla_{\mu}\mathcal{B}_{3}{}^{\rm c, m}$ can be removed since it is redundant.

Now we can expand the action up to second order. Then varying it with respect to $W_{i}{}^{\rm m}$ we get
\begin{equation}
    W_{i}{}^{\rm m} = -a n_{\rm m} V_{i} + \frac{n_{\rm m}}{\rho(n_{\rm m})_{,n_{\rm m}}} \mathcal{A}_{i}{}^{\rm m} \,.
\end{equation}
Substituting back the $W_{i}{}^{\rm m}$ to the action and varying the action with respect to the field $\mathcal{A}_{i}{}^{\rm m}$ we get
\begin{equation}
    \mathcal{A}_{i}{}^{\rm m} = a (V_{i} - \dot{\mathcal{B}}_{i}{}^{\rm m}) \rho(n_{\rm m})_{n_{\rm m}} \,,
\end{equation}
and varying with respect to the field $\mathcal{B}_{i}{}^{\rm m}$ we get
\begin{equation}
    \dot{\mathcal{A}}_{i}{}^{\rm m} = 0\,.
\end{equation}
This show that $\mathcal{A}_{i}{}^{\rm m}$ is constant that is 
\begin{equation}
    a (V_{i} - \dot{\mathcal{B}}_{i}{}^{\rm m}) \rho(n_{\rm m})_{n_{\rm m}} = \text{constant}\,.
\end{equation}

Now from the dark matter sector varying the action with respect to the field $W_{i}{}^{\rm c}$ we get
\begin{equation}
    \mathcal{A}_{i}{}^{\rm c} = 2 a n_{\rm c} V_{i} + 2 W_{i}{}^{\rm c} + \left( \frac{Y_{i}}{n_{\rm c}} - \frac{a \phi}{n_{\rm c}} V_{i} - \frac{\phi}{n_{\rm c}{}^{2}} W_{i}{}^{\rm c}  \right) f_{, \mathcal{Z}} + \frac{\phi}{n_{\rm c}} \frac{\dot{Y}_{i}}{a} f_{,\mathcal{E}} \,,
\end{equation}
where 
\begin{equation}
    Y_{i} \equiv E_{i} + a \phi V_{i} \,.
\end{equation}
Following similar procedure for the standard matter fluid as done before, we substitute $W_{i}{}^{\rm c}$ from the above constraint equation into the action. Then we can find the equation of motion for the field $\mathcal{A}_{i}{}^{\rm c}$
\begin{equation}
    \mathcal{A}_{i}{}^{\rm c} = 2 n_{\rm c} a V_{i} \rho(n_{\rm c})_{,n_{\rm c}} - 2 n_{\rm c} a \dot{\mathcal{B}}_{i}{}^{\rm c} + \left( \frac{Y_{i}}{n_{\rm c}} - \frac{a \phi}{n_{\rm c}} V_{i} + \frac{a \phi}{n_{\rm c}} \dot{\mathcal{B}}_{i}{}^{\rm c} \right) f_{, \mathcal{Z}} + \frac{\phi}{n_{\rm c}} \frac{\dot{Y}_{i}}{a} f_{, \mathcal{E}} \,.
\end{equation}
Now varying the action with respect to the field $\mathcal{B}_{i}{}^{c}$ we get
\begin{equation}
    \dot{\mathcal{A}}_{i}{}^{\rm c} = 0 \,.
\end{equation}
This shows that $\mathcal{A}_{i}{}^{\rm c}$ is constant. That is
\begin{equation}
    2 n_{\rm c} a V_{i} \rho(n_{\rm c})_{,n_{\rm c}} - 2 n_{\rm c} a \dot{\mathcal{B}}_{i}{}^{\rm c} + \left( \frac{Y_{i}}{n_{\rm c}} - \frac{a \phi}{n_{\rm c}} V_{i} + \frac{a \phi}{n_{\rm c}} \dot{\mathcal{B}}_{i}{}^{\rm c} \right) f_{, \mathcal{Z}} + \frac{\phi}{n_{\rm c}} \frac{\dot{Y}_{i}}{a} f_{, \mathcal{E}} = \text{constant}.
\end{equation}

By varying the action with respect to the field $V_{i}$ we get 
\begin{equation}
    \frac{1}{2} M_{\rm P}^{2} a^2 k^2 V_{i} + 2 a^{4}\tilde{n}^{2} \left(V_{i} - \delta\dot{\mathcal{B}_{i}{}^{\rm c}}\right) +  2 a^{4} n_{\rm m} \left(V_{i} - \delta\dot{\mathcal{B}}_{i}{}^{\rm m}\right) - \phi a^4  \left(V_{i} - \delta\dot{\mathcal{B}}_{i}{}^{\rm c}\right) f_{,\mathcal{Z}} + a^3 f_{,\mathcal{Z}} Y_{i} + a^{2} \phi \dot{Y}_{i} f_{, \mathcal{E}} = 0 \,.
\end{equation}
On substituting back the $V_{i}$, and taking the action in the small-scale limit $(k \rightarrow \infty)$ we can find the action contains
\begin{equation}
    S_{V}{}^{(2)} \simeq \sum_{i = 1, 2} \int {\rm d}t {\rm d}^{3} x a^2\left[ Q_{V} \left(  \frac{\dot{Y}_{i}^{2}}{a^2} - c_{V}^{2} \frac{k^2}{a^2} Y_{i}^{2} \right) \right] \,,
\end{equation}
where 
\begin{equation} \label{eq:Ghost_vector}
    Q_{V} = \frac{1}{2}\frac{\rho_{\rm c} + \phi \left( f_{,\mathcal{Z}} -\phi f_{,\mathcal{E}}  \right)}{\rho_{\rm c} + \phi \left( f_{,\mathcal{Z}} \right)} \,, \qquad c_{V}^{2} = \frac{1}{2 Q_{V}} \,.
\end{equation}
This shows that, there is no additional vector mode present, but the kinetic term is modified. Hence, the condition for no-ghost is given by,
\begin{equation}
    Q_{V} = \frac{1}{2}\frac{\rho_{\rm c} + \phi \left( f_{,\mathcal{Z}} -\phi f_{,\mathcal{E}}  \right)}{\rho_{\rm c} + \phi \left( f_{,\mathcal{Z}} \right)} > 0\,.
\end{equation}
This also ensures the avoidance of Laplacian instabilities for the propagating vector modes $Y_{i}$. Notice that we have started from the vector fields, whose $z$ component is zero. Hence, we have two vector modes that propagating in the $z$ direction.

\subsubsection{Scalar mode}

In this part, we discuss the linear stability of the scalar modes. We expand the action without choosing any gauge and make the field redefinition,
\begin{equation}
    \psi_{\phi} \equiv \phi \chi + \delta \phi_{s}\,,
\end{equation}
which remove the second derivative of the variable $\chi$. We also make the following field redefinition
\begin{eqnarray}
    \delta J_{\rm m} = \frac{N_{\rm m}}{a^3}\left( \frac{\rho}{n\rho_{,n}} \delta_{\rm m } - \alpha \right) \,, \label{eq:Jmredef}\\
    \delta n_{0} = \frac{N_{\rm c}}{a^3} \left( \frac{\rho_{c}}{2 \tilde{n} \rho_{c,\tilde{n}}} \delta_{\rm c} - \alpha \right) \,. \label{eq:n0redef}
\end{eqnarray}
Now we find the equations of motion for the field $j_{\rm m}$ and $n_{s}$ 
\begin{eqnarray}
    \delta l & = &  \rho_{,n_{\rm m}} \chi + \frac{\rho_{,n_{\rm m}}}{n_{\rm m}} j_{\rm m} \, , \\
    \delta K & = & \frac{N_{\rm c}}{a^3} \left( 2 \rho_{c,\tilde{n}} \chi  + f_{,\mathcal{E}} \frac{a^6}{N_{c}^{2} }\right) + \left( 2 \rho_{c,\tilde{n}} + f_{,\mathcal{Z}}  \frac{a^6}{N_{\rm c}{}^{2}} \right) n_{s} - \frac{a^3}{N_{\rm c}}f_{,\mathcal{Z}} \psi_{\phi} \nonumber \\
    & & - f_{,\mathcal{E}} \frac{a^3}{N_{\rm c}} \left( \phi \, \delta \phi + 2 \phi^2 \, \alpha  + \phi \, \frac{\dot{\psi}_{\phi}}{a} \right) \, .
\end{eqnarray}
We replace $\delta l$ and $\delta K$ with the above equations of motions and also make another field redefinition
\begin{eqnarray}
    j_{\rm m} & = & -\frac{N_{\rm m}}{a^3} \frac{a}{k^2}\theta_{\rm m} \,, \\
    n_{s} & = & - \frac{N_{\rm cmd}}{a^3} \frac{a}{k^2}\theta_{\rm c}\,.
\end{eqnarray}

After a few integration by parts, we get the second order action, which can be written as
\begin{eqnarray}
    S_{\rm Proca} & = & \int d^{4}x \,\, a^4 \left[ \chi  \left(-w_{1}\frac{k^2 \alpha}{a ^2} - w_{2} \frac{k^2 \delta \phi}{a^2 \phi} - y_{1} \frac{2 k^2
   \dot{\zeta} }{a }\right) +\delta \phi  \left( w_{6} \frac{k^2 \partial_{t}\left(E/a^2 \right) }{a} - w_{3} \frac{k^2 \dot{\psi}_{\phi} }{2 a ^3 \phi^2}+\frac{k^2 \psi_{\phi}(w_{2} + w_{6} \phi)}{2 a^2 \phi^2} - w_{6}\frac{3 \dot{\zeta} }{a }\right) \right. \nonumber \\
   & & \left. +\alpha  \left(\delta \phi  \left(- w_{3} \frac{k^2}{a^2 \phi} - \frac{3 H w_{1} - 2 w_{4}}{\phi}\right) + k^2 \left(w_{1}\frac{ \partial_{t}\left(E/a^2 \right)}{a }+ 2 y_{1} \zeta \right) - w_{3} \frac{k^2 \dot{\psi}_{\phi} }{a^3 \phi} + w_{6} \frac{k^2 \psi_{\phi}}{a ^2} - w_{1}\frac{3 \dot{\zeta} }{a }\right) \right. \nonumber \\
   & & + \alpha ^2 \left( w_{4} - w_{3} \frac{k^2}{a ^2}\right) + k^2 \left(\frac{2 M_{\rm P}^2 \partial_{t}\left(E/a^2 \right) \dot{\zeta} }{a ^2}+\zeta  \left( y_{2} \frac{\partial_{t}\left(E/a^2 \right) }{a} - y_{4} E/a^2 \right)+ y_{2} \frac{E/a^2 \dot{\zeta} }{a }+ y_{1} \zeta ^2\right) \nonumber \\
   & & \left. +\delta \phi^2 \left( w_{5} \frac{1 }{\phi^2} - w_{3} \frac{k^2}{4 a^2 \phi^2}\right) - w_{3} \frac{k^2 \dot{\psi}_{\phi} ^2}{4 a ^4 \phi
    ^2} + w_{7}\frac{k^2 \psi_{\phi} ^2}{2 a ^2} - \frac{3 M_{\rm P}^{2} \dot{\zeta} ^2}{a ^2} + y_{3} k^4 \left(-E/a^2 \right) + y_{3} 3 \zeta^2 \right] \,,
\end{eqnarray}
\begin{eqnarray}
    S_{\mathcal{E}} & = & \int d^{4}x \,\, a^4 \left[ f_{\mathcal{E}}\left\{ \chi  \left[ k^2 \left(-\frac{\delta \phi  \phi  }{a ^2}-\frac{\phi     \dot{\psi}_{\phi} }{a ^3}\right)-\frac{2 k^2 \alpha  \phi  ^2}{a ^2}\right] \right. \right. \nonumber \\
    & & +\alpha  \left[-\frac{2 \rho_{c}  \phi  ^2
   \dot{\delta}_{c} }{a  (P_{c} +\rho_{c} )}+k^2 \left(\frac{2 \phi  ^2 \partial_{t} \left(E/a^2\right) }{a }+\frac{2 \psi_{\phi}  \phi
    }{a ^2}\right)-\frac{6 \phi  ^2 \dot{\zeta} }{a }+\frac{6 \delta_{c}  H  \phi  ^2 (\delta \rho_{c} 
   P_{c} -\delta P_{c}  \rho_{c} )}{\delta \rho_{c}  (P_{c} +\rho_{c} )}\right] \nonumber  \\
   & & +\dot{\delta}_{c} 
   \left(-\frac{\delta \phi  \rho_{c}  \phi  }{a  (P_{c} +\rho_{c} )}-\frac{\rho_{c}  \phi   \dot{\psi}_{\phi} }{a ^2
   (P_{c} +\rho_{c} )}\right)+\delta_{c}  \left( H  \phi \frac{3 \dot{\psi}_{\phi}  (\delta \rho_{c} 
   P_{c} -\delta P_{c}  \rho_{c} )}{a  \delta \rho_{c}  (P_{c} +\rho_{c} )} +  H  \phi \frac{3 \delta \phi (\delta \rho_{c}  P_{c} -\delta P_{c}  \rho_{c} )}{\delta \rho_{c}  (P_{c} +\rho_{c} )}\right) \nonumber \\
   & & \left. +k^2 \left(  \phi  \frac{\delta \phi \partial_{t} \left(E/a^2 \right) }{a}+\frac{\delta \phi  \psi_{\phi} }{a^2} + \phi \frac{\partial_{t} \left(E/a^2\right) \dot{\psi}_{\phi} }{a^2} - \frac{H  \psi_{\phi}^2}{2 a ^2}\right) - \phi \frac{3 \delta\phi \dot{\zeta} }{a } - \phi \frac{3   \dot{\psi}_{\phi}  \dot{\zeta} }{a ^2}\right\} \nonumber \\
   & & \left. -\frac{k^2 \psi_{\phi} ^2 \dot{\phi}  f_{,\mathcal{Z E}}}{2
   a ^3} - \frac{k^2 \psi_{\phi} ^2 \phi \dot{\phi} f_{,\mathcal{Y E}}}{2 a ^3} \right] \,,
\end{eqnarray}
\begin{eqnarray}
    S_{\mathcal{Z}} & = & \int d^{4}x \,\, a^{4} \left[ f_{,\mathcal{Z}} \left\{\chi  \left(\frac{\rho_{c}  \phi   \dot{\delta}_{c} }{a (P_{c} +\rho_{c} )}+k^2 \left(-\frac{\phi   \partial_{t}\left( E/a^2 \right)  }{a }-\frac{\psi_{\phi} }{a ^2}\right)+\frac{3 \phi \dot{\zeta} }{a }+\frac{3 \delta_{c}  H  \phi   (\delta P_{c}  \rho_{c} -\delta \rho_{c} 
   P_{c} )}{\delta \rho_{c}  (P_{c} +\rho_{c} )}\right) \right. \right. \nonumber \\
   & &  + \phi \frac{k^2 \chi ^2}{2 a ^2}+\theta_{c} \left(- 3aH \phi \frac{ \delta_{c} (\delta P_{c}  \rho_{c} -\delta \rho_{c}  P_{c} )}{k^2 \delta \rho_{c} (P_{c} +\rho_{c} )} - \rho_{c}\phi\frac{\dot{\delta}_{c} }{k^2 (P_{c} +\rho_{c} )}+\phi \partial_{t}\left( E/a^2 \right) - 3 \phi \frac{\dot{\zeta} }{k^2}\right) + \delta_{c}  \left(\frac{\rho_{c}  \dot{\psi}_{\phi} }{a (P_{c} +\rho_{c} )}+\frac{3 H  \psi_{\phi}  \rho_{c} }{P_{c} +\rho_{c} }\right) \nonumber \\
   & & \left. +k^2 \left(-\frac{(E/a^2) 
   \dot{\psi}_{\phi}}{a }+\frac{\psi_{\phi} ^2}{2 a ^2 \phi  }-3 (E/a^2)  H  \psi_{\phi} \right)+\frac{3 \zeta \dot{\psi}_{\phi} }{a }-\alpha  \delta\phi -\frac{1}{2} \phi \alpha^2 -\frac{\delta \phi ^2}{2 \phi  }+9 H \psi_{\phi}  \zeta - \phi \frac{\theta_{c} ^2}{2 k^2}\right\} \nonumber \\
   & & + f_{,\mathcal{ZZ}} \left\{ \frac{\dot{\phi}}{a} \frac{\delta_{c}  \psi_{\phi}  \rho_{c}}{(P_{c} +\rho_{c} )}- \frac{\dot{\phi}}{a} k^2 (E/a^2) \psi_{\phi} + 3 \frac{\dot{\phi}}{a} \psi_{\phi}\zeta + \phi \, \alpha  \delta \phi +\frac{1}{2}\phi^2 \, \alpha^2 +\frac{1}{2} \delta \phi^2\right\} \nonumber \\
   & &  \left. + f_{,\mathcal{YZ}} \left\{ \phi \frac{\dot{\phi}}{a} \frac{\delta_{c} \psi_{\phi}  \rho_{c} }{(P_{c} +\rho_{c} )} - \phi \frac{\dot{\phi}}{a} k^2 (E/a^2)  \psi_{\phi} + 3 \phi \frac{\dot{\phi}}{a} \psi_{\phi} \zeta + 2   \phi^2 \, \alpha  \delta \phi+ \phi^3 \, \alpha^2 +\phi \, \delta \phi^2 \right\} \right] \,,
\end{eqnarray}
\begin{eqnarray}
    S_{I} & = & \sum_{I = c,b,r} \int d^{4}x a^4 \left[ \chi  \left(\frac{\rho_{I}  \dot{\delta}_{I} }{a }-\frac{k^2 \partial_{t}\left( E/a^2 \right)  (P_{I} +\rho_{I} )}{a }+\frac{3
   (P_{I} +\rho_{I} ) \dot{\zeta} }{a }+3 \delta_{I}  H  \left(\frac{\delta P_{I} 
   \rho_{I} }{\delta \rho_{I} }-P_{I} \right)\right)+\frac{k^2 \chi ^2 (P_{I} +\rho_{I} )}{2
   a ^2} \right. \nonumber   \\
   & & +\theta_{I}  \left(\frac{3 a  \delta_{I}  H  (\delta \rho_{I}  P_{I} -\delta P_{I}  \rho_{I} )}{k^2
   \delta \rho_{I} }-\frac{\rho_{I}  \dot{\delta}_{I} }{k^2}+\partial_{t}\left( E/a^2 \right)  (P_{I} +\rho_{I} )-\frac{3
   (P_{I} +\rho_{I} ) \dot{\zeta} }{k^2}\right)-\alpha  \delta_{I}  \rho_{I} -\frac{\delta_{I} ^2
   \delta P_{I}  \rho_{I} ^2}{2 \delta \rho_{I}  (P_{I} +\rho_{I} )} \nonumber  \\
   & & \left. +k^2 (E/a^2)  \zeta 
   (-P_{I} -\rho_{I} )+\frac{1}{2} k^4 (E/a^2) ^2 (-P_{I} -\rho_{I} )+\frac{\theta_{I} ^2
   (-P_{I} -\rho_{I} )}{2 k^2}+\frac{3}{2} \zeta ^2 (P_{I} +\rho_{I} ) + \rho_{r} \delta_{r} \sigma_{r}  \right] \,.
\end{eqnarray}

The coefficients $w_{i}$ and $y_{i}$ are given as 
\begin{eqnarray}
    w_{1} & = & 2M_{\rm P}^{2} H - \phi^3 G_{,\mathcal{Y}} \,, \\
    w_{2} & = & 2 M_{\rm P}^{2} H - w_{1}\,, \\
    w_{3} & = & - 2\phi^2 \,, \\
    w_{4} & = & -3 M_{\rm P}^{2} H^2 - \frac{3}{2} H \phi^3 \left( G_{3,\mathcal{Y}} - \phi^2 G_{3,\mathcal{YY}} \right) + \frac{1}{2}\phi^4 f_{,\mathcal{YY}} \,, \\
    w_{5} & = & -\frac{3}{2} H (w_{1} + w_{2}) + w_{4} \,, \\
    w_{6} & = & -\phi^2 G_{3,\mathcal{Y}} \,, \\
    w_{7} & = & \frac{\dot{\phi}}{a} \frac{w_{2}}{\phi^3} \,, \\
    y_{1} & = & M_{\rm P}^{2}/a^2 \,, \\
    y_{2} & = & - M_{\rm P}^{2} H - \phi G_{3} \,, \\
    y_{3} & = & M_{\rm P}^{2} \frac{\dot{H}}{a} \frac{1}{2}\phi^2 w_{7} \,, \\
    y_{4} & = & -3 H y_{2} + 3 M_{\rm P}^{2} \frac{\dot{H}}{a} + \frac{\dot{\phi}}{a} G_{3} \,.
\end{eqnarray}
We choose a flat gauge for simplicity, i.e. we keep the variables $\zeta = 0$, and $E = 0$. Then we have the variables, $\alpha$, $\chi$, from the gravity sector $\delta_{\rm m}$, $\delta_{\rm c}$, $\delta l_{\rm m}$, $\delta K$, $\delta j_{\rm m}$ and $n_{s}$ in the matter and dark matter sector. Finally, we have $\delta \phi$ and $\psi_{\phi}$ in the dark energy sector.

Now we integrate out the variables $\delta j_{\rm m}$, $n_{s}$, $\delta l_{\rm m}$, $\delta K$, $\chi$, $\delta \phi$, and $\alpha$. We are remaining with the variables $\delta_{\rm m}$, $\delta_{\rm c}$, and $\psi_{\phi}$, and notice that all of them are dynamical. After integration by part we can write the action in the form
\begin{equation}
\label{eq:can_action}
    \int {\rm d}^3x {\rm d}t \, a^4 \left[ \dot{\mathcal{U}}^{T}\, \mathbf{K}\, \dot{\mathcal{U}} -\frac{k^2}{a^2} \mathcal{U}^{T}\, \mathbf{G} \, \mathcal{U} - \mathcal{U}^{T}\, \mathbf{M} \, \mathcal{U} - \frac{k}{a} \mathcal{U}^{T}\, \mathbf{B} \, \dot{\mathcal{U}} \right] \,,
\end{equation}
where $\mathbf{K}$, $\mathbf{G}$, $\mathbf{M}$, and $\mathbf{B}$ are 4$\times$4 matrices and the vector $\mathcal{U}^{T}$ is given by 
\begin{equation}
    \mathcal{U}^{T} \equiv (\psi_{\phi}, \delta_{c}/k, \delta_{b}/k, \delta_{r}/k)\,.
\end{equation}
Here, we consider the matter components baryon and radiation.

In the high $k$ limit ($k\rightarrow \infty$) we have 
\begin{eqnarray}
    \mathbf{K}_{11} & = & \frac{1}{2(w_{1} - 2 w_{2})^2 \phi^2} \left[ 3 M_{\rm P}^2 H^2 w_{1}^{2} + 4 M_{\rm P}^{4} H^2 w_{4} + 2 M_{\rm P}^{4} H^2 \phi \left( - f_{,\mathcal{Z}} + 
    \phi f_{,\mathcal{ZZ}} + 2 \phi^2 f_{,\mathcal{YZ}} \right) \right] \,, \label{eq:K11} \\ 
    \mathbf{K}_{22} & = & \frac{a^2 \rho_{c}^2\left[ P_{\rm c} + \rho_{c} + \phi \left( - \phi f_{,\mathcal{E}} + f_{,\mathcal{Z}}\right) \right]}{ (\rho_{c} + P_{\rm c})^2} \,, \qquad  \mathbf{K}_{33}  =  \frac{a^2 \rho_{r}^2}{2 (\rho_{r} + P_{r})}\,, \qquad \mathbf{K}_{44} = \frac{a^2 \rho_{b }^2}{2 (\rho_{b} + P_{b} )}\,, \label{eq:K223344}
\end{eqnarray}
and $\mathbf{G}$ in the leading order 
\begin{eqnarray}
    \mathbf{G}_{11} & = & \mathcal{G} + H \mu + \frac{\dot{\mu}}{a} - \frac{w_{2}{}^2\sum_{I} \big(\rho_{I} + P_{I} \big)}{2 \phi^2 \big( w_{1} - 2 w_{2} \big)^2} - \frac{M_{\rm P}^{4} H^2 f_{, \mathcal{Z}}}{\phi^2 \big( w_{1} - 2 w_{2} \big)^2} + \frac{2 M_{\rm P}^{4} f_{, \mathcal{E} } \big( -2 \phi G_{3, \mathcal{Y}} + f_{,\mathcal{E}} \big)}{\big( w_{1} - 2 w_{2} \big)^2}\,, \\ 
    \mathbf{G}_{22} & = & \frac{a^2 c_{c}^{2} \rho_{c}^2}{2 (\rho_{c} + P_{c})}\,, \qquad  \mathbf{G}_{33}  =  \frac{a^2 c_{r}^{2} \rho_{r}^2}{2 (\rho_{r} + P_{r})}\,, \qquad \mathbf{G}_{44} = \frac{a^2 c_{b}^{2} \rho_{b }^2}{2 (\rho_{b} + P_{b} )}\,,
\end{eqnarray}
where, 
\begin{equation}
    \mathcal{G} \equiv - \frac{4 M_{\rm P}^{4} H^{2} w_{2}{}^{2}}{\phi^2 \big( w_{1} - 2 w_{2} \big)^2 w_{3}} - \frac{w_{2}\dot{\phi}}{2 a \phi^3}\,, \qquad \text{and} \qquad \mu \equiv \frac{M_{P}^{2} H w_{2}}{\big(w_{1} - 2 w_{2} \big) \phi^2}\,.
\end{equation}

The leading order term for the anti-symmetric matrix is
\begin{equation}
\label{eq:antisymmB}
    \mathbf{B}_{12} = - \mathbf{B}_{21} = - \frac{M_{\rm P}^{2} a H \rho_{c} ( \phi^2 G_{3,\mathcal{Y}} f_{,\mathcal{E}} - \phi f_{,\mathcal{E}}^{2} + f_{,\mathcal{Z}})}{(\rho_{c} + P_{c}) (2 M_{\rm P}^{2} H - \phi^2 G_{3,\mathcal{Y}}) } \,.
\end{equation}

We assume that the fluid obeys the weak energy condition $\rho_{\rm I} + P_{\rm I} > 0$, for ($I= c,b,r$), which follows the standard ghost conditions for baryons and radiation.
\begin{equation}
    \mathbf{K}_{33} >0 \,, \qquad \mathbf{K}_{44} > 0 \,.
\end{equation}
Then we have the following ghost conditions for the fields $\psi_{\phi}$ and $\delta_{c}$. 
\begin{eqnarray}
    q_{s} & \equiv &  3 w_{1}^{2} + 4 M_{\rm P}^{2} w_{4} + 2 M_{\rm P}^{2} \phi \left( - f_{,\mathcal{Z}} + \phi f_{,\mathcal{ZZ}} + 2 \phi^2 f_{,\mathcal{YZ}} \right) > 0 \,, \\
    q_{c} & \equiv & 1 + \frac{\phi \left( - \phi f_{,\mathcal{E}}^2 + f_{,\mathcal{Z}}\right)}{(\rho_{c} + P_{c})} > 0 \,. \label{eq:qc}
\end{eqnarray}

The propagation of the speed for baryons and radiation is not modified by the $\mathbf{B}$ matrix. Then their speeds of propagation are 
\begin{eqnarray}
    c_{b}^{2} = \frac{\mathbf{G}_{33}}{\mathbf{K}_{33}} \,, \qquad c_{r}^{2} = \frac{\mathbf{G}_{44}}{\mathbf{K}_{44}} \,.
\end{eqnarray}

However, the other two dynamical fields $\psi_{\phi}$ and $\delta_{c}$ are affected by the anti-symmetric metric $\mathbf{B}$ Eq.~(\ref{eq:antisymmB}). Now taking the equations of motion from the action Eq.~(\ref{eq:can_action}), on assuming a solution 
\begin{equation}
    \psi_{\phi} = \bar{\psi}_{\phi} e^{i(\omega t - kx)} \,, \qquad \delta_{c} = \bar{\delta}_{c} e^{i(\omega t - kx)} \,, 
\end{equation}
we have
\begin{eqnarray}
    \omega^2 \bar{\psi}_{\phi} - \hat{c}_{s}^{2} \frac{k^2}{a^2}\bar{\psi}_{\phi} - i \omega \frac{k}{a} \frac{\mathbf{B}_{12}}{\mathbf{K}_{11}}\bar{\delta}_{c}& \simeq & 0 \,, \label{eq:EQdeltac2} \\
    \omega^2 \bar{\delta}_{c} - \hat{c}_{c}^{2} \frac{k^2}{a^2}\bar{\delta}_{c} - i \omega \frac{k}{a} \frac{\mathbf{B}_{21}}{\mathbf{K}_{11}} \bar{\psi}_{\phi}  & \simeq & 0 \,,  \label{eq:EQpsiphi2}
\end{eqnarray}
where 
\begin{eqnarray}
    \hat{c}_{s}^{2} & = & \frac{2(w_{1} - 2 w_{2})^2 \phi^2}{M_{\rm P}^{2} H^{2}q_{S}}  \left[ \mathcal{G} + H \mu + \frac{\dot{\mu}}{a} - \frac{w_{2}{}^2\sum_{I} \big(\rho_{I} + P_{I} \big)}{2 \phi^2 \big( w_{1} - 2 w_{2} \big)^2} - \frac{M_{\rm P}^{4} H^2 f_{, \mathcal{Z}}}{\phi^2 \big( w_{1} - 2 w_{2} \big)^2} + \frac{2 M_{\rm P}^{4} f_{, \mathcal{E} } \big( -2 \phi G_{3, \mathcal{Y}} + f_{,\mathcal{E}} \big)}{\big( w_{1} - 2 w_{2} \big)^2} \right] \,,\\
    \hat{c}_{c}^{2} & = & \frac{\mathbf{G}_{22}}{\mathbf{K}_{22}} = \frac{c_{c}^{2}}{q_{c}} \,. 
\end{eqnarray}

Now we know that $c_{c}^{2}=0$, then it follows that $\hat{c}_{c}^{2} =0$. The equation of motion Eq.~(\ref{eq:EQpsiphi2}) will be reduced to 
\begin{equation}
     \omega^2 \bar{\delta}_{c} - i \omega \frac{k}{a} \frac{\mathbf{B}_{21}}{\mathbf{K}_{11}} \bar{\psi}_{\phi}   \simeq 0 \,.
\end{equation}
There are two solutions,
\begin{eqnarray}
    \omega & = & 0 \,, \label{eq:Solbaranch1}\\
    \omega \bar{\delta}_{c} & = & i \frac{k}{a} \frac{\mathbf{B}_{21}}{\mathbf{K}_{11}} \bar{\psi}_{\phi} \label{eq:Solbaranch2} \,.
\end{eqnarray}
The first branch of the solution corresponds to 
\begin{equation}
    c_{c}^{2} = \frac{\omega^2 a^2}{k^2} = 0 \,.
\end{equation}

The other branch corresponds to the propagation of the longitudinal mode $\delta\phi$. By substituting the second branch Eq.~(\ref{eq:Solbaranch2}) in to the equation of motion Eq.~(\ref{eq:EQdeltac2}) we get
\begin{equation}
    \frac{k^2}{a^2}c_{s}^{2} = \omega^2 \,, \qquad  c_{s}^2 = \hat{c}_{s}^{2} + \Delta c_{s}^{2} \,,
\end{equation}
where 
\begin{equation}
\label{eq:Deltacs2}
    \Delta c_{s}^{2} = \frac{1}{q_{s}q_{c}} \left[ \frac{M_{\rm P}^{2} a H \rho_{c} ( \phi^2 G_{3,\mathcal{Y}} f_{,\mathcal{E}} - \phi f_{,\mathcal{E}}^{2} + f_{,\mathcal{Z}})}{(\rho_{c} + P_{c}) (2 M_{\rm P}^{2} H - \phi^2 G_{3,\mathcal{Y}}) } \right]^2.
\end{equation}

\section{Linear perturbation equations of motion}\label{sec:linear_perturbation_eom}
Now let us find the equations of motion for the linear perturbation. The second order Lagrangian we are considering have the following fields,
\begin{equation}
    ( \alpha, \quad \chi, \quad \zeta, \quad E, \quad \delta \phi, \quad \psi_{\phi}, \quad \delta_{\rm m}, \quad \theta_{\rm m}, \quad \delta_{\rm c}, \quad \theta_{\rm c} ).
\end{equation}
Notice that we have not made any choice of gauge. Instead, after varying for each field above, we make a choice of the Newtonian Gauge, 
\begin{equation}
    \alpha = \Psi, \qquad \zeta = - \Phi, \qquad \chi = 0, \qquad \text{and} \qquad E = 0.
\end{equation}

\subsection{Matter sector}
Next, let us look at the equations of motion for the matter sector:
\begin{eqnarray}
        E_{\theta_{\rm m}} & \equiv &  \dot{\delta}_{\rm m} + 3aH \left( \frac{\delta P_{\rm m}}{\delta \rho_{\rm m}} - \frac{P_{\rm m}}{\rho_{\rm m}} \right) \delta + \left(1 + \frac{P_{\rm m}}{\rho_{\rm m}} \right)\theta_{m} - 3 \left( 1 + \frac{P_{\rm m}}{\rho_{\rm m}}  \right) \dot{\Phi} =0\,, \\
        E_{\delta_{\rm m}} & \equiv & \dot{\theta}_{\rm m} + aH \left( 1 - 3 \frac{\delta P_{\rm m}}{\delta \rho_{\rm m}} \right) \theta_{\rm m} - k^2 \left( \frac{\delta P_{\rm m}/\delta \rho_{\rm m}}{1+ P_{\rm m}/\rho_{\rm m}} \right)\delta_{\rm m} - k^2 \Psi + k^2 \sigma = 0 \,.
\end{eqnarray}

The above equations are the equations of motion of the standard matter fluids in the system we are considering, in particular, radiation and baryon. However, for baryon, we have $P_{\rm m}=0$ and $\delta P_{\rm m}=0$. The field $\sigma$ denotes the shear of the fluid, for example, for photons. 

\subsection{Dark matter sector}
Now we find the following equations of motion for the dark matter fluid, together with the interaction with the vector dark energy. Here, we can see the implications of the interaction terms we have considered.
\begin{eqnarray}
    E_{\theta_{\rm c}} & \equiv & \dot{\delta}_{\rm c} + \theta_{\rm c} - 3 \dot{\Phi} = 0 \,, \\
    E_{\delta_{\rm c}} & \equiv & \dot{\theta}_{\rm c} + \left[ \frac{1}{\rho_{c} + \phi f_{,\mathcal{Z}}} \left\{ aH (\rho_{c} + 4 \phi f_{,\mathcal{Z}} ) + \dot{\phi} \big( f_{,\mathcal{Z}} + \phi (f_{\mathcal{ZZ}} + \phi f_{,\mathcal{ZY}}) \big) \right\} \right] \theta_{\rm c} \nonumber \\
    & & + \left[ \frac{ k^2 }{\rho_{c} + \phi f_{,\mathcal{Z}}} \right] \left( f_{,\mathcal{Z}} \frac{\dot{\psi}_{\phi}}{a} - \rho_{c} \Psi \right) + \left[ \frac{ k^2 }{a (\rho_{c} + \phi f_{,\mathcal{Z}})} \big(3aH f_{,\mathcal{Z}} + \dot{\phi}(f_{,\mathcal{ZZ}} + \phi f_{,\mathcal{ZY}} ) \big) \right] \psi_{\phi} \nonumber \\
    & & + \left[ \frac{ k^2 f_{,\mathcal{E}}}{\rho_{c} + \phi f_{,\mathcal{Z}}} \right] \left( \phi\frac{\delta\dot{\phi}}{a} + 2 \phi^2 \frac{ \dot{\Psi}}{a} + \phi\frac{\ddot{\psi}_{\phi}}{a^2} + \frac{(2aH \phi + \dot{\phi})}{a} \frac{\delta \dot{\phi}_{s}}{a} + 2 \phi \frac{(3aH \phi + 2 \dot{\phi})}{a}  \Psi + \frac{(3aH \phi + \dot{\phi})}{a} \delta \phi \right) \nonumber \\
    & & + \left[ \frac{ k^2 f_{,\mathcal{EZ}}}{\rho_{c} + \phi f_{,\mathcal{Z}}} \right] \left( \frac{\phi \dot{\phi}}{a} (\delta\phi + 2 \Psi + \frac{ \dot{\psi}_{\phi}}{a}) \right) + \left[ \frac{k^2 f_{,\mathcal{EY}}}{\rho_{c} + \phi f_{,\mathcal{Z}}} \right] \left( \frac{\phi^2 \dot{\phi}}{a} \big( \delta\phi + 2 \phi \Psi + \frac{ \dot{\psi}_{\phi}}{a} \big) \right) = 0 \,.
\end{eqnarray}
Notice that the continuity equation for the cold dark matter density perturbation is not affected by the interaction. However, the evolution equation for the velocity divergence is modified significantly due to the interaction that we are considering. This shows that all the interaction terms that we are considering contribute to the momentum transfer. Also, note that there is a term in $ \ddot{\psi}_{\phi}$, proportional to $f_{\mathcal{E}}$, which we will replace with the equation for the field $\psi_{\phi}$.

\subsection{Gravity sector}
Finally, we will show the equations of motion for the field that comes in the gravity sector, including the vector dark energy. Let us first look at the equations of motion by varying the Proca fields.
\begin{eqnarray}
    E_{\delta \phi} & \equiv & \frac{k^2}{a^2} \left(\delta \phi + 2 \phi \Psi - \phi G_{3,\mathcal{Y}} \psi_{\phi}+ \frac{ \dot{\psi}_{\phi}}{a} \right) + G_{3,\mathcal{Y}} \left( 3H \phi^2 + \frac{3 \phi^3 \dot{H}}{\dot{\phi}} \right) \Psi - 3 \phi^2 G_{3,\mathcal{Y}} \frac{\dot{\Phi}}{a} - 3 \phi^2 G_{3,\mathcal{Y}} \frac{\dot{H}}{\dot{\phi}} \delta\phi \nonumber \\
    & & + f_{,\mathcal{E}} \left( \frac{k^2}{a^2} \psi_{\phi} + \frac{\phi}{a} \theta_{\rm c}  \right) = 0 \, , \label{eq:EQdphi} \\
    E_{\psi_{\phi}} & \equiv & f_{,\mathcal{E}}\left[-\phi\frac{   \ddot{\delta}_{c} }{a ^2}+k^2
   \left(-\frac{\delta \phi }{a ^2}+ H \frac{\psi_{\phi}}{a^2}- 2 \phi \frac{\Psi}{a^2}\right)+2 H \phi \frac{\theta_{c}}{a} + 3 \phi \frac{\ddot{\Phi}}{a^2} + \dot{\phi} \frac{\theta_{c}}{a ^2}\right]  \nonumber \\
    & & +f_{,\mathcal{ZE}} \left[\dot{\phi}\frac{k^2 \psi_{\phi}}{a^3} +  \phi   \dot{\phi} \frac{\theta_{c} }{a ^2}\right]+f_{,\mathcal{YE}} \left[ \phi   \dot{\phi} \frac{k^2\psi_{\phi} }{a ^3} + \phi^2 \dot{\phi} \frac{\theta_{c}  }{a ^2}\right] + f_{,\mathcal{Z}}\left[-\frac{k^2 \psi_{\phi} }{a ^2 \phi  }-\frac{\theta_{c} }{a }\right] \nonumber \\ 
    & & +k^2 \left[\frac{\delta \dot{\phi} }{a ^3}+\delta \phi  \left(\frac{\phi   G_{3,\mathcal{Y}}}{a^2}+\frac{H }{a ^2}\right)+\Psi  \left(\frac{\phi^2 G_{3,\mathcal{Y}}}{a^2}+\frac{2 H \phi}{a^2}+\frac{2 \dot{\phi} }{a ^3}\right)+\frac{\psi_{\phi}  \dot{\phi} G_{3,\mathcal{Y}}}{a ^3}+\frac{2
   \phi \dot{\Psi} }{a ^3}+\frac{\ddot{\psi}_{\phi} }{a ^4}\right] = 0 \,. \label{eq:EQPsiphi}
\end{eqnarray}

For the other fields in the metric, we have the following equations of motion
\begin{eqnarray}
    E_{\alpha} & \equiv & M_{\rm P}^{2} \left[ 2 \frac{k^2}{a^2} \Phi + 6 H \Psi + 6 \frac{H}{a} \dot{\Phi} \right] + \left( 3H \phi^2 G_{3,\mathcal{Y}} + \frac{3 \phi^2 \dot{H} G_{3,\mathcal{Y}}}{\dot{\phi}} - 2 \frac{k^2}{a^2} \phi \right)\, \delta\phi + \frac{k^2}{a^2}\phi^2 G_{3,\mathcal{Y}} \psi_{\phi} -2 \phi \frac{k^2}{a^2} \frac{ \dot{\psi}_{\phi}}{a}  \nonumber \\
    & & + 3 \phi^3 G_{3,\mathcal{Y}} \frac{\dot{\Phi}}{a} + \left( 6 H \phi^3 G_{3,\mathcal{Y}} + \frac{3 \phi^4 \dot{H} G_{3,\mathcal{Y}}}{\dot{\phi}} - 4 \frac{k^2}{a^2} \phi^2 \right)\, \Psi - 2 \phi f_{,\mathcal{E}} \left( \frac{k^2}{a^2} \psi_{\phi} + \frac{\phi}{a} \theta_{\rm c} \right)  + \delta \rho_{\rm m} + \delta \rho_{\rm c}= 0\,, \label{eq:EQalpha}  \\
    E_{\chi} & \equiv & M_{\rm P}^{2} k^2 \left( 2H \Psi + 2 \frac{\dot{\Phi}}{a}  \right) - a (\rho_{\rm m} + P_{\rm m}) \theta_{\rm m} - a \left( \rho_{c} + \phi f_{,\mathcal{Z}} \right) \theta_{\rm c} + k^2 f_{,\mathcal{Z}}  \psi_{\phi} +k^2 \phi^2 G_{3,\mathcal{Y}} \delta \phi + k^2 \phi^3 G_{3,\mathcal{Y}} \Psi \nonumber \\
    &  & - k^2 f_{,\mathcal{E}} \left( \phi \, \delta\phi + 2 \phi^2 \Psi + \phi  \frac{ \dot{\psi}_{\phi}}{a} \right) = 0 \,, \label{eq:EQchi} \\
    E_{\zeta} & \equiv & 2 M_{\rm P}^{2} \left[\frac{\ddot{\Phi}}{a^2} + 2 H \frac{\dot{\Phi}}{a} + H \frac{\dot{\Psi}}{a} + \left( 3 H^2 + \frac{2\dot{H}}{a} \right) \Psi + \frac{k^2}{a^2} \left( \frac{\Phi}{3} - \frac{\Psi}{3} \right) \right] + G_{3,\mathcal{Y}} \left( \phi^2 \frac{\delta\dot{\phi}}{a} + \phi^3 \frac{\dot{\Psi}}{a} \right) \nonumber \\
    & & +  \left[ G_{3,\mathcal{Y}} \left( 3H \phi^2 + 2 \phi \frac{\dot{\phi}}{a} \right) + \phi^3 G_{3,\mathcal{YY}} \frac{\dot{\phi}}{a} \right] \left( \frac{\delta}{\phi} + \Psi \right) -  \delta P_{\rm m} = 0 \,. \label{eq:EQzeta}
\end{eqnarray}
In the equation $E_{\alpha}$ we have used the equation $E_{\theta_{\rm c}}$ to remove the term in $\dot{\delta}_{\rm c}$, which is proportional to $f_{,\mathcal{E}}$. We have used the evolution of the velocity divergence to arrive at the equation $E_{\zeta}$.

From the linear combination of $E_{\zeta}$ and $E_{E}$, i.e. 
\begin{equation}
    E_{\rm shear} \equiv \frac{3}{k^2} E_{E} + E_{\zeta} = 0\,,
\end{equation}
we can find the shear equation
\begin{equation}
    E_{\rm shear} \equiv M_{\rm P}^2 k^2 (\Psi - \Phi) + \frac{3}{2} a^2 (P_{r} + \rho_{r}) \sigma_{r} = 0 \,. \label{eq:EQshear}
\end{equation}

Now we have all the equations of motion given in the Newtonian gauge. 

\section{Quasi Static Approximation and effective gravitational couplings} \label{sec:QSA}

In this section, we look at the equations of motion in the quasi-static limit. Then in the following subsection, we find the effective gravitational coupling for the dark matter and the baryon.

Let us start from the matter equation. 
\begin{eqnarray}
    E_{\theta_{\rm b}} & \equiv &  \dot{\delta}_{\rm b}  + \theta_{\rm b} = 0 \,, \label{eq:EthetamQSA}\\
    E_{\delta_{\rm b}} & \equiv & \dot{\theta}_{\rm b} + aH \theta_{\rm b}  - k^2 \Psi = 0,  \, \label{eq:EdeltamQSA}
\end{eqnarray}
where we have considered that the contribution of matter is from baryon, i.e. $P_{\rm b}=0$. Taking a time derivative of the (\ref{eq:EthetamQSA}) and substituting (\ref{eq:EdeltamQSA}), we find the second order equation for the baryonic over density 
\begin{equation}\label{eq:EQdeltam2nd}
    \ddot{\delta}_{\rm b} + aH \dot{\delta}_{\rm b} + k^2 \Psi = 0  \, .
\end{equation}

Following the same procedure for the dark matter sector we have the continuity equation
\begin{eqnarray}
    E_{\theta_{\rm c}} & \equiv &  \dot{\delta}_{\rm c}  + \theta_{\rm c} = 0 \, , \label{eq:EQthetadmQSA}
\end{eqnarray}
which is not affected by the interaction. We 
notice that in the evolution equation of the velocity divergence $\theta_{c}$, there is a term proportional to $\ddot{\psi}_{\phi}$, which is replaced with equation of motion for the field $\psi_{\phi}$ Eq.~(\ref{eq:EQPsiphi}). After doing this replacement, the equation becomes 

\begin{eqnarray}
    & & \ddot{\delta}_{\rm c} \left(1 - \frac{\phi^2 f_{,\mathcal{E}}}{\rho_{c} + \phi f_{,\mathcal{Z}}} \right) - \dot{\psi}_{\phi} \left[ \frac{k^2 \left(\phi  ^2 f_{,\mathcal{E}} G_{3,\mathcal{Y}} -\phi f_{,\mathcal{E}}^2+f_{,\mathcal{Z}} \right)}{a  \left(\phi f_{,\mathcal{Z}}+\rho_{c} \right)} \right] + \Psi \frac{k^2}{\rho_{c} + \phi f_{,\mathcal{Z}}} \left[ -\rho_{c} +\phi^3 G_{3,\mathcal{Z}} f_{,\mathcal{E}} \right] \nonumber \\
    & & - \psi_{\phi} \frac{k^2}{\rho_{c} + \phi f_{,\mathcal{Z}}} \left[ a  \left( H  \left(2 \phi  ^2 f_{,\mathcal{E}} G_{3,\mathcal{Y}} - 3 \phi  
   f_{,\mathcal{E}}^2 + 3 f_{,\mathcal{Z}} + f_{,\mathcal{E}}  \left(2 \phi  ^2 f_{,\mathcal{E}} G_{3,\mathcal{Y}} - \phi f_{,\mathcal{E}}{}^2+f_{,\mathcal{Z}} + \phi^3 \left(-G_{3,\mathcal{Y}}^2\right)\right)\right) \right. \right. \nonumber \\
   & & \left. + \dot{\phi}  \left(\phi  ^3 f_{,\mathcal{YE}} G_{3,\mathcal{Y}} + \phi^2 f_{,\mathcal{EZ}} G_{3,\mathcal{Y}} - 2 \phi f_{,\mathcal{E}} 
   \left(f_{,\mathcal{EZ}} +\phi  
   f_{,\mathcal{YE}}\right)+\phi   f_{,\mathcal{YZ}} - f_{,\mathcal{E}}^2+f_{,\mathcal{ZZ}}\right) \right] \nonumber \\
   & & - \dot{\delta}_{c} \frac{k^2}{\rho_{c} + \phi f_{,\mathcal{Z}}} \left[a  \left(\phi   f_{,\mathcal{E}}\left(\phi  ^2 f_{,\mathcal{E}}
   G_{3,\mathcal{X}} - \phi   f_{,\mathcal{E}} ^2+f_{,\mathcal{Z}} \right) + H  \left(4 \phi  \left(f_{,\mathcal{Z}} - \phi f_{,\mathcal{E}}^2\right)+\rho_{c} \right)\right) \right. \nonumber \\
   & & \left. - \dot{\phi} \left(-2 \phi^3 f_{,\mathcal{E}}   f_{,\mathcal{YE}} + \phi^2 \left(f_{,\mathcal{YZ}} - 2f_{,\mathcal{E}} f_{,\mathcal{ZE}}\right)+\phi \left(f_{,\mathcal{ZZ}} - 2 f_{,\mathcal{E}} ^2\right)+f_{,\mathcal{Z}}\right) \right] =0 \,.
\end{eqnarray}
In the above expression we have used the fact that, from Eq.~(\ref{eq:EQshear})
\begin{equation}
    \Psi = \Phi \, ,
\end{equation}
since baryons do not have any shear. Eq.~(\ref{eq:EQdphi}) in the quasi-static limit gives 
\begin{equation}\label{eq:EQdohiQSA}
    \delta \phi \simeq - 2\phi \Psi + G_{3,\mathcal{Y}} \phi \psi_{\phi} - \frac{\dot{\psi}_{\phi}}{a} - f_{,\mathcal{E}} \left(\psi_{\phi} + \frac{a \phi \theta_{\rm c}}{k^2} \right) \,,
\end{equation}
$\theta_{\rm c}$ can be expressed in terms of $\dot{\delta}_{\rm c}$ using Eq.~(\ref{eq:EQthetadmQSA}).

Now we want to find the solution of the fields $\Psi  (\simeq \Phi)$ and $\psi_{\phi}$ in the quasi-static limit. We start from the Eq.~(\ref{eq:EQalpha}), $E_{\alpha}$, in the quasi-static limit 
\begin{equation}\label{eq:EQalphaQSA}
    \frac{k^2}{a^2} \left( - 2M_{\rm P}^2 \Phi + 2 \phi \delta\phi + 4 \phi^2 \Psi + 2 \frac{\dot{\phi}}{a} \delta \dot{\psi}_{\phi} - \phi^2 G_{3,\mathcal{Y}}  \right) + \left(2 \frac{k^2}{a^2} \phi \psi_{\phi} + 2 \frac{\phi^2}{a} \theta_{\rm c} \right) f_{,\mathcal{E}} - \rho_{\rm b} \delta_{\rm  b} - \rho_{\rm c} \delta_{\rm c}  \simeq 0 \,.
\end{equation}
Substituting the Eq.~(\ref{eq:EQdohiQSA}) to the above equation Eq.~(\ref{eq:EQalphaQSA}) we get 
\begin{equation}\label{eq:EQalphaQSA2}
    -2 M_{\rm P}^2\frac{k^2}{a^2} \Phi + \frac{k^2}{a^2}\phi^2G_{3,\mathcal{Y}} \psi_{\phi} - \rho_{\rm b} \delta_{\rm b} - \rho_{c} \delta_{\rm c} \simeq 0 \,. 
\end{equation}

We take a time derivative of the Eq.~(\ref{eq:EQalphaQSA2})
\begin{eqnarray}\label{eq:DEQalphaQSA2Dt}
    & & 4 M_{\rm P}^2\frac{k^2 H \Phi }{a} + 3 a H \rho_{\rm b} \delta_{\rm b} + 3 a H \rho_{c} \delta_{\rm c} - \rho_{\rm b} \dot{\delta}_{\rm b} - \rho_{c} \dot{\delta}_{\rm c} - \frac{k^2}2 M_{\rm P}^2{a^2}\dot{\Phi} + 2H\frac{k^2}{a^2} \phi^2 G_{3,\mathcal{Y}} \dot{\psi}_{\phi} \nonumber \\ 
    & & \frac{k^2}{a^2} \left( - 2aH \phi^2 G_{3,\mathcal{Y}} + 2 \phi \dot{\phi} G_{3,\mathcal{Y}} + \phi^3 \dot{\phi} G_{3,\mathcal{YY}} \right) \psi_{\phi}\simeq 0 \,.
\end{eqnarray}
From the quasi-static limit of  the Eq.~(\ref{eq:EQchi}), $E_{\chi}$, we get 
\begin{eqnarray} \label{eq:EQchiQSA} 
    k^2 \left(f_{,\mathcal{E}} \left(-\frac{\phi \dot{\psi}_{\phi}}{a} - \delta \phi \phi  -2 \Psi \phi  ^2\right)- \psi_{\phi} f_{,\mathcal{Z}} + \delta \phi \phi^2 \phi f_{,\mathcal{Z}} - a  \rho_{c} \right) + k^2 M_{\rm P}^{2} \left(\frac{2 \dot{\Phi} }{a }+2
   H  \Psi\right)-a  \rho_{b}  \theta_{b}  \simeq 0\,. 
\end{eqnarray}
From the linear combination of Eq.~(\ref{eq:DEQalphaQSA2Dt}) and Eq.~(\ref{eq:EQchiQSA}), and replacing $\Psi (\simeq \Phi)$ with Eq.~(\ref{eq:EQalphaQSA2}), $\theta_{\rm b}$ with Eq.~(\ref{eq:EthetamQSA}), and $\theta_{\rm c}$ Eq.~(\ref{eq:EQthetadmQSA}) respectively we get 
\begin{eqnarray} \label{eq:EQchiQSA2}
    & & \dot{\delta}_{c}  \left(\frac{\phi  ^3 f_{,\mathcal{E}}G_{3,\mathcal{Y}}
    }{a } - \frac{\phi^2 f_{,\mathcal{E}}^2}{a }+\frac{\phi f_{,\mathcal{Z}}}{a}\right) \nonumber \\
    & &   +\psi_{\phi} \left(-\frac{2 k^2 \phi^2 f_{,\mathcal{E}}
   G_{3,\mathcal{Y}}}{a^2}+\frac{k^2 \phi f_{,\mathcal{E}}^2}{a^2}-\frac{k^2
   f_{,\mathcal{Z}}}{a ^2}+\frac{k^2 \phi  ^3 \dot{\phi}  G_{3,\mathcal{Y}} }{a ^3}
   +\frac{k^2 H  \phi  ^2 G_{3,\mathcal{Y}}}{a ^2}-\frac{k^2 \phi  ^5 G_{3,\mathcal{Y}}^2}{2 M_{\rm P}^{2} a ^2}+\frac{2 k^2 \phi   \dot{\phi}  G_{3,\mathcal{Y}}}{a ^3}+\frac{k^2 \phi  ^3
   G_{3,\mathcal{Y}}^2}{a ^2}\right) \nonumber  \\
   & & +\frac{\delta_{c} \rho_{c} \phi^3 G_{3,\mathcal{Y}}}{2 M_{\rm P}^{2}}+\frac{\delta_{b} \rho_{b}\phi  ^3 G_{3,\mathcal{Y}}}{2 M_{\rm P}^{2}} \simeq 0  \,.
\end{eqnarray}
Hence, we can use Eq.~(\ref{eq:EQalphaQSA2}) and Eq.~(\ref{eq:EQchiQSA2}) to solve for $\Psi (\simeq \Phi)$ and $\psi_{\phi}$, respectively. 

\subsection{Effective gravitational coupling}
The second order equation of motion for the baryonic over density Eq.~(\ref{eq:EQdeltam2nd}) substituting for $\Psi (\simeq \Phi)$ from Eq.~(\ref{eq:EQalphaQSA2}), we get
\begin{equation}
    \ddot{\delta}_{\rm b} + aH \dot{\delta}_{\rm b} + a \Gamma_{\rm bc} \dot{\delta}_{\rm c} - \frac{3}{2} \frac{H^2}{G} a^2 (G_{\rm bb}\Omega_{\rm b} \delta_{\rm b} + G_{\rm bc}\Omega_{\rm c} \delta_{\rm c} ) = 0 \,,
\end{equation}
where 
\begin{eqnarray}
    & & G_{\rm  bb}/G = G_{\rm bc}/G \nonumber \\
    & & 1 - \frac{a  \phi^5 G_{3,\mathcal{Y}}^2}{a  \left(4 M_{\rm P}^{2} \phi^2 f_{,\mathcal{E}} G_{3,\mathcal{Y}} - 2 M_{\rm P}^{2} \phi f_{,\mathcal{E}}^2+2 M_{\rm P}^{2} f_{,\mathcal{Z}} - 2 M_{\rm P}^{2} H  \phi^2 G_{3,\mathcal{Y}} - 2 M_{\rm P}^{2} \phi^3 G_{3,\mathcal{Y}}^2 + \phi^5 G_{3,\mathcal{Y}}^2\right)-2 M_{\rm P}^{2} \phi   \dot{\phi}  \left(\phi^2 G_{3,\mathcal{YY}} + 2 G_{3,\mathcal{Y}}\right)}\,, \nonumber \\
\end{eqnarray}
and 
\begin{equation}
    \Gamma_{bc} \equiv \frac{a  \phi  ^3 G_{3,\mathcal{Y}} \left(\phi^2 f_{\mathcal{E}} G_{3,\mathcal{Y}} - \phi   f_{\mathcal{E}}^2+f_{,\mathcal{Z}}\right)}{a  \left(4 M_{\rm P}^{2} \phi^2 f_{\mathcal{E}} G_{3,\mathcal{Y}} - 2 M_{\rm P}^{2} \phi   f_{\mathcal{E}}^2 + 2 M_{\rm P}^{2} f_{,\mathcal{Z}} - 2 M_{\rm P}^{2} H \phi^2 G_{3,\mathcal{Y}} - 2 M_{\rm P}^{2} \phi^3 G_{3,\mathcal{Y}}^2+\phi^5 G_{3,\mathcal{Y}}^2\right)-2 M_{\rm P}^{2} \phi   \dot{\phi} \left(\phi^2 G_{3,\mathcal{YY}} + 2 G_{3,\mathcal{Y}}\right)}
\end{equation}
In the above equation, we have replaced $\rho_{\rm b}$ with $(3H^2 M_{\rm P}^2)\Omega_{\rm b}$ and $\rho_{c}$ with $(3H^2 M_{\rm P}^2)\Omega_{\rm c}$. It is interesting to notice that when $G_{3,\mathcal{Y}} \rightarrow 0$ and $G_{3,\mathcal{YY}} \rightarrow 0$, $\Gamma_{\rm bc} = 0 $ and $G_{\rm bb} = G_{\rm bc} = G$. 

Now for the equation of motion for the density evolution of dark matter on replacing $\Psi (\simeq \Phi)$ by Eq.~(\ref{eq:EQalphaQSA2}), $\psi_{\phi}$ by Eq.~(\ref{eq:EQchiQSA2}), and $\dot{\psi}_{\phi}$ by time derivative of Eq.~(\ref{eq:EQchiQSA2}), we get
\begin{eqnarray}
\label{eq:delta_c2ndorder}
    \ddot{\delta}_{\rm c} + a \Gamma_{\rm cc} \, \dot{\delta}_{\rm c} + a  \Gamma_{\rm cb} \, \dot{\delta}_{\rm b} - \frac{3}{2} \frac{H^2}{G} a^2 (G_{\rm cb}\Omega_{\rm b} \delta_{\rm b} + G_{\rm cc}\Omega_{\rm c} \delta_{\rm c} ) = 0\,,
\end{eqnarray}
The expression for $G_{\rm cc} (= G_{\rm cb})$ and $\Gamma_{\rm cc}$ and $\Gamma_{\rm cb}$ are given in the appendix.

\section{Concrete model}\label{sec:con_model}
Until now the interaction term that we considered in the action is completely general, that is, the interaction is introduced in the Lagrangian as a general function $f(\mathcal{Y,Z,E})$. In order to study phenomenological consequences of these interactions, we consider a concrete model given by 
\begin{equation}\label{eq:concrete_model}
    f(\mathcal{Y,Z,E}) \equiv b_{2}\mathcal{Y}^{p_{2}} + \beta \mathcal{Z} + \gamma \mathcal{E} \,, \qquad G_{3}(\mathcal{Y}) \equiv g_{3} \mathcal{Y}^{p_{3}}\,.
\end{equation}
\subsection{Background and ghost conditions}
Introducing the density parameters $\Omega_{\rm DE}$, $\Omega_{r}$, $\Omega_{c}$, and $\Omega_{b}$, the first Friedmann equation Eq.~(\ref{eq:1stFE}) can be rewritten as 
\begin{equation}
    1 - \Omega_{r} - \Omega_{c} -\Omega_{b} - \Omega_{\rm DE} = 0 \,,
\end{equation}
where 
\begin{equation}
    \Omega_{i} \equiv \frac{\rho_{i}}{3 M_{\rm P}^{2} H^{2}} \,.
\end{equation}
From the first Friedmann equation we get the energy density of the dark energy applying the concrete model 
\begin{equation}
    \rho_{\rm DE} = - 2^{p_{2}} b_{2} \phi^{2 p_{2}} - \beta \phi \,.
\end{equation}
The constraint equation of background, Eq.~(\ref{eq:EQphi}), can be written as 
\begin{equation}
    \beta + 2^{1-p_{2}}b_{2} p_{2} \phi^{2p_{2}-1}  + 3 \times 2^{1-p_{3}} g_{3} p_{3} H \phi^{2 p_{3}} = 0 \,.
\end{equation}
In this concrete model, the relation between the Hubble expansion $H$ and $\phi$ is given by
\begin{equation}
    \phi^{p} H = \lambda = \text{constant}\,,
\end{equation}
where the relation between the powers are given by
\begin{equation}
    p = 2 p_{3} - 2 p_{2} + 1 \,, \quad p_{2} = \frac{1}{2} \,.
\end{equation}

Now, taking the time derivative of Eq.~(\ref{eq:EQphi}), we get 
\begin{equation}
    \phi  ^2 \left( \dot{\phi}  f_{,\mathcal{YY}}+3 \dot{H}  G_{3,\mathcal{Y}}\right)+\phi   \dot{\phi}\left(2 f_{,\mathcal{YZ}} +3 H  G_{3,\mathcal{Y}} \right)+ \dot{\phi} 
   f_{,\mathcal{ZZ}} - \frac{ \dot{\phi}  f_{,\mathcal{Z}} }{\phi  }+3 H  \phi
    ^3  \dot{\phi}  G_{3,\mathcal{YY}}  = 0 \,.
\end{equation}
In the concrete model given by Eq.~(\ref{eq:concrete_model}), we get 
\begin{equation}
    \dot{\phi} = - \frac{s a H (-3 + 3 \Omega_{\rm DE} - \Omega_{\rm r}) \phi}{(1 + s \Omega_{\rm DE})} \,.
\end{equation}
The above relation can be used to replace $\dot{\phi}$ in the expressions.

Finally, with this concrete model, we can write
\begin{equation}
    \phi = \phi_{0} \left( \frac{\Omega_{\rm DE}}{\Omega_{{\rm DE}, 0}} \right)^{s/(1+s)}\,, \qquad H = H_{0} \left( \frac{\Omega_{{\rm DE}, 0}}{\Omega_{\rm DE}} \right)^{1/[2(1+s)]} \,.
\end{equation}

\begin{figure}
    \centering
    \includegraphics[scale=1.2]{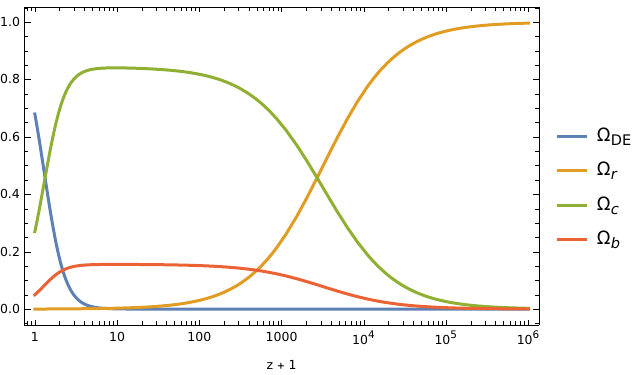}
    \caption{Evolution of $\Omega_{\rm DE}$, $\Omega_{r}$, $\Omega_{c}$, $\Omega_{b}$. Here we chose the value of $s=0.2$. As for the initial condition we have chosen $\Omega_{{\rm DE},0} = 0.68$, $\Omega_{r,0} = 10^{-4}$ $\Omega_{c, 0} = 0.27 $, and $\Omega_{b, 0} = 0.05$, at the redshift $z= 0$. }
    \label{fig:Omega_i_evo}
\end{figure}

The evolution of the density parameters $\Omega_{\rm DE}$, $\Omega_{r}$ and $\Omega_{\rm c}$ can be found as 
\begin{eqnarray}
    \Omega^{\prime}_{\rm DE}  & = & \frac{(1+s)\Omega_{\rm DE} (3 - 3\Omega_{\rm DE} + \Omega_{r})}{1 + s \Omega_{\rm DE}} \,, \\
    \Omega^{\prime}_{r}  & = & \frac{(1+(3+4s)) \Omega_{\rm DM} - \Omega_{r})\Omega_{r}}{1 + s \Omega_{\rm DE}} \,, \\
    \Omega^{\prime}_{c}  & = & - \frac{(3 (1+ s) \Omega_{\rm DE} - \Omega_{r})\Omega_{c}}{1 + s \Omega_{\rm DE}} \,,
\end{eqnarray}
where the prime denotes the derivative with respect to the variable $\mathcal{N} \equiv \ln (a)$, $s = 1/(2p)$. The evolution of the background density parameters are shown in Figure \ref{fig:Omega_i_evo}. The above evolution equation is first derived in~\cite{DeFelice:2016yws}.

\begin{figure}
    \centering
    \includegraphics[scale=1.2]{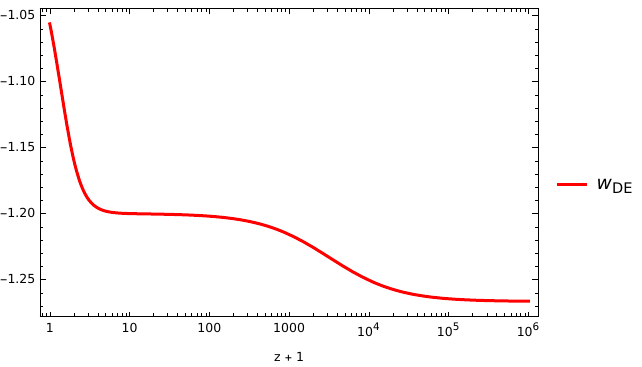}
    \caption{Behaviour of $w_{\rm DE}$. Here also we chose the value of $s=0.2$, and we use same initial condition to solve the density parameter differential equations, $\Omega_{{\rm DE},0} = 0.68$, $\Omega_{r,0} = 10^{-4}$ $\Omega_{c, 0} = 0.27 $, and $\Omega_{b, 0} = 0.05$, at the redshift $z= 0$. }
    \label{fig:w_de}
\end{figure}

From the background equation of motion, we can write the equation of state for the dark energy as 
\begin{equation}
    w_{\rm DE} = - \frac{3 + 3 s + s \Omega_{r}}{3 + 3 s \Omega_{\rm DE}} \,.
\end{equation}
An example of the behavior of $w_{\rm DE}$ is shown in Figure \ref{fig:w_de}.

The ghost condition for the dynamical field $\psi_{\phi}$, Eq.~(\ref{eq:K11}), becomes
\begin{equation}\label{eq:qs}
    \mathbf{K}_{11} = \frac{3 M_{\rm P}^{2} H^{2} \Omega_{\rm DE} (1+ s \Omega_{\rm DE})}{s(\Omega_{\rm DE} -2 )^2 \phi^2} \,,
\end{equation}
and the speed of propagation of the longitudinal mode $c_{s}^{2} = \hat{c}_{s}^{2} +\Delta c_{s}^{2}$, where, $\hat{c}_{2}^{2}$ becomes
\begin{eqnarray}
     \hat{c}_{s}^{2} & = & -\frac{s \left(-s \Omega_{\rm DE} ^2 \left((\phi/M_{\rm P})  ^2-2\right)+2 \Omega_{\rm DE}  \left((s-2) (\phi/ M_{\rm P})^2+1\right)+(\Omega_{\rm r} +5) (\phi/M_{\rm P})  ^2\right)}{3 (\phi / M_{\rm P} )^2 (s \Omega_{\rm DE} +1)^2} \nonumber \\
    & & -\frac{2 r_{\beta} s}{s \Omega_{\rm DE} +1}+\frac{2 r_{\gamma} s \phi /M_{\rm P} }{s \Omega_{\rm DE} +1}-\frac{4 s\sqrt{r_{\gamma}\Omega_{\rm DE} }}{\sqrt{3} \sqrt{\phi/M_{\rm P}  } (s \Omega_{\rm DE} +1)} \,,
\end{eqnarray}
where we have redefined the parameters
\begin{equation}
    \Omega_{\gamma} = \frac{\gamma^2 \phi^2}{3 M_{\rm P}^{2} H^2} \,, \qquad \Omega_{\beta} = \frac{\beta \phi}{3 M_{\rm P}^{2} H^2} \,, \qquad r_{\beta} = \frac{\Omega_{\beta}}{\Omega_{\rm DE}} = \frac{- 2 \beta}{\sqrt{2} b_{2} + 2 \beta} \,, \qquad r_{\gamma}/M_{\rm P} = \left(\frac{\Omega_{\gamma}}{\Omega_{\rm DE}} \right) \frac{1}{\phi} =  \frac{-2 \gamma^2 }{\sqrt{2} b_{2} + 2 \beta} \,,
\end{equation}
where the parameters $r_{\beta}$ and $r_{\gamma}$ are dimensionless. The quantity $\Delta c_{s}^{2}$ becomes,
\begin{eqnarray}
    \Delta c_{s}^{2} = \frac{2 s \Omega_{\rm DE}  \left[\sqrt{3 r_{\gamma} \Omega_{\rm DE}}  +3 \sqrt{\phi / M_{\rm P} } (r_{\beta}-r_{\gamma} \phi/M_{\rm P})\right]^2}{9 \phi /M_{\rm P} (s \Omega_{\rm DE} +1) \big[\Omega_{\rm c} +\Omega_{\rm DE}  (r_{\beta}-r_{\gamma} \phi/M_{\rm P}) \big]} \,.
\end{eqnarray}
The ghost condition for the cold dark matter content, Eq.~(\ref{eq:qc}), in the concrete model is given by 
\begin{equation}
     q_{c} = 1 + \frac{\beta \phi}{3 M_{\rm P}^{2} H^{2} \Omega_{c}} - \frac{\gamma^2 \phi^2}{3 M_{\rm P}^{2} H^{2} \Omega_{c}}  = 1 + r_{\beta} \frac{\Omega_{\rm DE}}{\Omega_{c}} - r_{\gamma}\frac{\Omega_{\rm DE}}{\Omega_{c}} \frac{\phi}{M_{\rm P}} = 1 + \frac{\Omega_{\rm DE}}{\Omega_{c}} \left( r_{\beta} - r_{\gamma} \frac{\phi}{M_{\rm P}} \right) \,.
\end{equation}
For the vector modes, we can write the no-ghost condition Eq.~(\ref{eq:Ghost_vector}) after choosing the concrete model as
\begin{equation}
    Q_{V} = \frac{1}{2} \left( 1 - \frac{  r_{\gamma} \Omega_{\rm DE} \phi }{M_{\rm P} \left[ \Omega_{\rm c} + r_{\beta} \Omega_{\rm DE} \right]}  \right) > 0\,.
\end{equation}


In the early epoch that is in the asymptotic past, $\Omega_{\rm DE} \rightarrow 0$, the ghost condition for the scalar mode
\begin{equation}
    q_{s} = 3 M_{\rm P}^4 H ^2 \Omega_{\rm DE}  (s \Omega_{\rm DE} +1)/s > 0 \,,
\end{equation}
implies that the parameter $s$ needs to satisfy the condition $s >0$. In the de Sitter fixed point, the avoidance of ghost also requires $s >0$.

The ghost condition for the cold dark matter in the radiation and the matter domination epoch becomes trivial $q_{c} = 1$. In the de Sitter epoch as well as the region when $\Omega_{\rm c} \rightarrow 0$, for the avoidance of ghost instability, we need the condition
\begin{equation} \label{eq:rbeta_rgamma_con}
    r_{\beta} > r_{\gamma} \phi_{\rm dS} \,,
\end{equation}
where $\phi_{\rm dS} = M_{\rm P} \phi =$ const. at the de Sitter fixed point.

For the vector mode, in the radiation and the matter domination, $Q_{V} = 1/2$. In the de Sitter epoch, the no-ghost condition becomes 
\begin{equation}
    Q_{V} = \frac{1}{2} \left(1 - \frac{r_{\gamma} \phi_{\rm dS}}{r_{\beta}} \right) \,.
\end{equation}
then the condition Eq.~(\ref{eq:rbeta_rgamma_con}), ensures that in the de Sitter dominant epoch, there is no ghost instability for the vector modes. 

For the speed of propagation of the longitudinal mode in the radiation domination, matter domination and the de Sitter epochs, the expression $c_{s}^{2}$ becomes
\begin{align}
    (c_{s}^{2})_{\rm ra} &= 2 s (1 - r_{\beta} )\,, \label{eq:cs2Ra}\\
    (c_{s}^{2})_{\rm ma} &= \frac{1}{3} s (5  - 6 r_{\beta}   ) \,, \label{eq:cs2Ma}\\ 
    (c_{s}^{2})_{\rm dS} & = - \frac{s \left(-r_{\beta} \phi_{\rm dS}^2+r_{\gamma} \phi_{\rm dS}^3-2 r_{\beta}\right)}{3 (s+1) \phi_{\rm dS}^2 (r_{\beta} - r_{\gamma} \phi_{\rm dS})} \,. \label{eq:cs2dS}
\end{align}
The quantity $\Delta c_{s}^{2}$ in the de Sitter fixed point becomes 
\begin{equation}
    (\Delta c_{s}^{2})_{\rm dS} = \frac{2 s \left(3 \sqrt{\phi_{\rm dS}} (r_{\beta}-r_{\gamma} \phi_{\rm dS} )+\sqrt{3 r_{\gamma}}\right)^2}{9 (s+1) \phi_{\rm dS} (r_{\beta}-r_{\gamma} \phi_{\rm dS} )} \,,
\end{equation}
whose contribution is zero in radiation domination and matter domination. 

For the avoidance of Laplacian instability, we need Eq.~(\ref{eq:cs2Ra}), Eq.~(\ref{eq:cs2Ma}), and Eq.~(\ref{eq:cs2dS}) to be positive in radiation, matter and de Sitter dominant epochs, respectively. Hence, we need the following condition to be satisfied
\begin{equation}
   0 < r_{\beta} < 5/6 \,,
\end{equation}
which will also ensure the condition
\begin{equation}
    \left(-r_{\beta} \phi_{\rm dS}^2+r_{\gamma} \phi_{\rm dS}^3-2 r_{\beta}\right) < 0 \,,
\end{equation}
as long as $r_{\beta}$ is positive.

\begin{figure}
    \centering
    \includegraphics[scale=1.2]{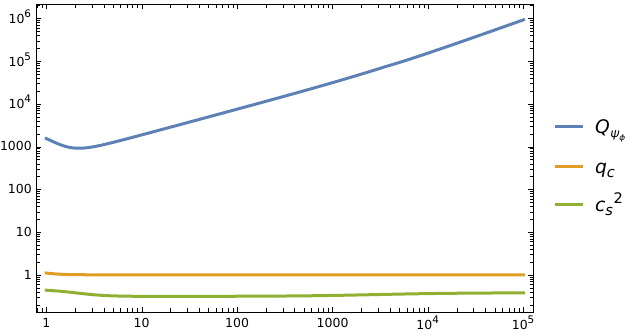}
    \caption{The behaviour of $Q_{\psi_{\phi}} \equiv \mathbf{K}_{11}M_{\rm P} /\phi  H^{2}$, $q_{c}$, and $c_{s}{}^{2}$. We chose the following values of the parameter $r_{\gamma} = 0.02$, $r_{\beta}= 0.05$, $s = 0.2$, $\phi_{\rm dS} = 0.489 $. }
    \label{fig:qs_and_cs2}
\end{figure}

\begin{figure}
    \centering
    \includegraphics[scale=1.2]{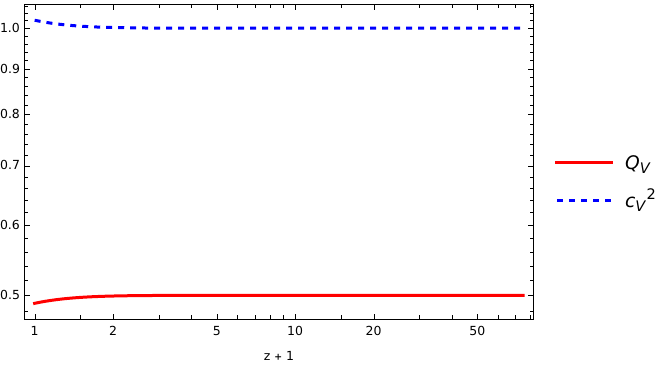}
    \caption{The behaviour of $Q_{V}$, and $c_{V}^{2}$. We chose the following values of the parameter $r_{\gamma} = 0.02$, and $r_{\beta} = 0.05$. }
    \label{fig:Qcs2_vector}
\end{figure}

With these parameter redefinition, we show an example of the behavior of the ghost conditions and the speed of propagation of the scalar mode in Figure \ref{fig:qs_and_cs2}.
The behaviours of $Q_{V}$ and the propagating speed $c_{V}^{2}$ are shown in Figure \ref{fig:Qcs2_vector}. 


\subsection{Dynamics of perturbation}
\begin{figure}
    \centering
    \includegraphics[scale=1.2]{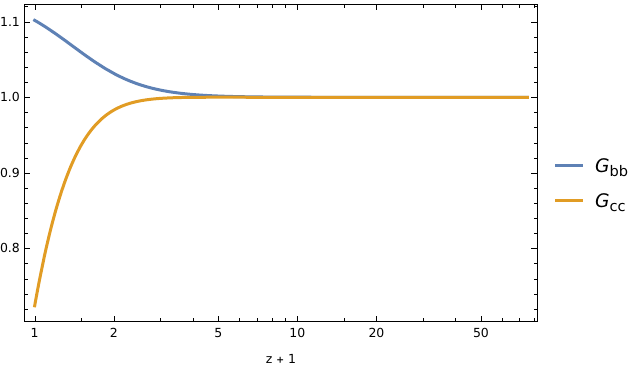}
    \caption{Behaviour of $G_{\rm cb}$, $G_{\rm cc}$. We chose the following values of the parameter $r_{\gamma} = 0.02$, $r_{\beta} = 0.05$, $s = 0.2$, $p = 5$, $\Omega_{{\rm DE},0} = 0.68$, $\phi_{\rm dS} = 0.489 $.}
    \label{fig:figureGandGamma}
\end{figure}

It has been shown that $\mathcal{Z}$ interaction can suppress the effective Newton constant, $G_{\rm eff}$. We have also seen that the new interaction that we have studied in this work can also modify $G_{\rm eff}$. Now we want to see the phenomenological effect of the interaction $\mathcal{E}$. The behavior of these quantities in the concrete model is shown Figure \ref{fig:figureGandGamma}. We can see that there is a suppression in $G_{\rm cc}$. 

\begin{figure}
    \centering
    \includegraphics[scale=1.5]{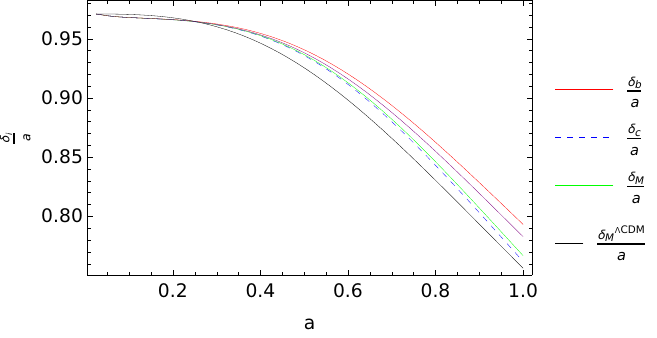}
    \caption{This is the result of numerically solving the equations Eq.~(\ref{eq:EQdeltac2}) and Eq.~(\ref{eq:delta_c2ndorder}). The red line shows the evolution of the baryon overdensity $\delta_{b}$. The dashed blue line shows the evolution of the cold dark matter overdensity $\delta_{\rm c}$. The green line shows the evolution of the total matter overdensity $\delta_{\rm M}$. The purple line shows the evolution of $\delta_{m}$ with interacting Proca background, keeping the over density equation for $\Lambda$CDM. The black line, show the evolution of the matter overdensity with $\Lambda$CDM background and perturbation.  It is clear that the evolution of the cold dark matter and total matter overdensities are suppressed compared with the baryon overdensity. Here, we chose the following values of the parameter $r_{\gamma} = 0.02$, $r_{\beta} = 0.05$, $s = 0.2$, $\Omega_{{\rm DE},0} = 0.68$, $\phi_{\rm dS} = 0.489$.}
    \label{fig:delta_evolution}
\end{figure}

We study the evolution of the baryon and cold dark matter overdensities through the evolution of the universe. This is shown in the Figure~\ref{fig:delta_evolution}.  In the figure, we also plot the perturbation equation for the matter with the background given by the interacting Proca model. 
The total matter evolution given by
\begin{equation}
    \delta_{\rm M} = \frac{\Omega_{b} \delta_{b}}{(\Omega_{b} + \Omega_{c})} + \frac{\Omega_{c} \delta_{c}}{(\Omega_{b} + \Omega_{c})} \,,
\end{equation}
is also shown in the figure. To solve the second order equations of motion, we convert the time variable to $\mathcal{N} = \ln{a}$. Then we give the initial conditions $\delta_{c} = \delta_{c}^{\prime} = 0.024$, and $\delta_{b} = \delta_{b}^{\prime} = 0.024$, at $\mathcal{N} = -3.7$.

In Figure~\ref{fig:delta_evolution}, we also plotted the evolution of the total matter density in the $\Lambda$CDM model. We noticed that,  $\delta_{M}$ in this model, is enhanced in comparison with the $\Lambda$CDM model with the same $\Omega_{\rm DE, 0}$. This is due to the modification in the background cosmology. To understand this behaviour better, we also show in Figure~\ref{fig:delta_evolution}, the purple line, which corresponds to the overdensity $\delta_{m}$ in the interacting Proca model in the background without modifications to the effective Newton constants. The modification in the background changes the coefficient of the fiction term in the second order matter overdensity equation, which causes the enhancement with respect to the total matter density in $\Lambda$CDM (the black line in the figure). The total matter overdensity in this model is suppressed due to the suppression of the effective Newton constant for cold dark matter (the green line in the figure), but this suppression is not sufficient to reduce the growth of matter overdensity in comparison with $\Lambda$CDM. 


\section{Conclusion}\label{sec:conclu}
In this work, we have constructed a theory of interacting dark energy and dark matter fluid (Section~\ref{sec:int_fluid_des}). The dark energy is described by a massive vector field, and the dark matter is described by a perfect fluid. We found a new interaction, $\mathcal{E} =A^{\alpha}F_{\alpha \beta}u^{\beta}$, which have not been explored. On top of this interaction, we also find the interactions that break the parity symmetry. On the basis of the parity-preserving interaction terms, we construct a general Lagrangian in Section~\ref{sec:interating_lagrangian}. Then, we introduced a toy model to understand the consequence of newly introduced interactions, where the dark energy action is a cubic generalised Proca theory. 

Then we studied the background evolution and linear perturbation equations of motion in Section~\ref{sec:cosmology}. We also derived the no ghost condition for the tensor, vector and scalar modes. Due to the term $\nabla_{\mu}K n^{\mu}$ in the dark matter action, this model does not modify the continuity equation, but it modifies the Euler equation. This indicates that the interaction introduces only a momentum transfer. Then, in Section~\ref{sec:linear_perturbation_eom}, we derived the linear equations of motion in Newtonian gauge.

To understand the phenomenological features of the theory, we study the evolution of baryon and cold dark matter in the high $k$ limit in Section~\ref{sec:QSA}. We show that the evolution of cold dark matter is modified as a result of the interaction. In particular, the effective gravitational coupling to the cold dark matter $G_{\rm cc}$ is modified. In the limit of $G_{3} \rightarrow 0$, we do not have any modification. For the evolution of baryons, we show that there are modifications similar to that of cold dark matter, particularly effective gravitational coupling $G_{\rm bb}$ is modified. 

We also introduced a simple concrete model for the free function $f$ introduced in the Lagrangian in Section~\ref{sec:con_model}, and studied the ghost conditions and speed of propagation of the dark energy scalar mode and vector modes. We also showed the behaviour of the modified effective gravitational couplings $G_{\rm bb}$ and $G_{\rm cc}$. We showed that the effective gravitational coupling $G_{\rm cc}$ was reduced. However, the effective gravitational coupling to baryons $G_{\rm bb}$ increases. The difference in the behaviour of the effective gravitational coupling to baryon and dark matter, when dark energy interacts only with dark matter, is a generic feature of interacting dark energy models as these models violate the equivalence principle~\cite{Kesden:2006vz, Liu:2023mwx, DeFelice:2020icf}. We numerically solved the evolution equation of baryon and cold dark matter over-densities and showed that the evolution of the cold dark matter was suppressed due to the interaction, as a result the total matter overdensity $\delta_{M}$ is also suppressed. To understand whether this model can address the $S_{8}$ tension or not, we also plotted the evolution of the total matter density in the $\Lambda$CDM model. We noticed that,  $\delta_{M}$ in this model, is enhanced in comparison with the $\Lambda$CDM model with the same $\Omega_{\rm DE, 0}$. This is due to the modification in the background cosmology. To understand this behaviour better, we also show in Figure~\ref{fig:delta_evolution}, the purple line, which corresponds to the overdensity $\delta_{m}$ in the interacting Proca model in the background without modifications to the effective Newton constants. The modification in the background changes the coefficient of the fiction term in the second order matter overdensity equation, which causes the enhancement with respect to the total matter density in $\Lambda$CDM (the black line in the figure). The total matter overdensity in this model is suppressed due to the suppression of the effective Newton constant for cold dark matter (the green line in the figure), but this suppression is not sufficient. The observable $\sigma_{8} (S_{8})$ is determined by the total matter overdensity $\delta_{\rm M}$ as gravitational lensing is determined by the gravitational potential, which is sourced by the total matter density. Since the evolution of $\delta_{M}$ of this model is not suppressed with respect to that of $\Lambda$CDM, as shown in Figure~\ref{fig:delta_evolution}, to address the $S_{8}$ tension, we need to extend the concrete model to see if this model can further suppress the growth of structure to address the $S_{8}$ tension. We note that Ref. \cite{DeFelice:2020icf} considered an extension of the $\mathcal{Z}$ interaction to the simplest one considered in this paper and showed that it is possible to suppress the growth compared with $\Lambda$CDM. We also need to constrain $\Omega_{\rm DE, 0}$ from the background observations to make a fair comparison with the $\Lambda$CDM prediction. We leave this for future work. We stress that all the formulae that are necessary to check the stability conditions and the modifications to the effective Newton constants were derived for the general Lagrangian given by Eq.~(\ref{eq:action_example}) in this paper.  


There are several future directions based on the current work. It is also interesting to study the implications of the interaction between the vector dark energy and dark matter with parity-violating interacting terms. 
It will also be interesting to see if this model can address the Hubble and large-scale structure $S_{8}$ tension simultaneously by extending the concrete model, modelling this as the early interacting dark sector. Another interesting direction is to study non-linear cosmology, which is important in the light of the upcoming cosmological surveys like Euclid~\cite{EUCLID:2011zbd}, Square Kilometer Array~\cite{Weltman:2018zrl}, and Rubin observatory's LSST~\cite{LSSTDarkEnergyScience:2012kar}. It is also interesting to find new interactions with vector dark energy models with a disformal coupling to dark matter~\cite{Achour:2021pla,Takahashi:2022ctx}.

\begin{acknowledgments}
M.C.P. acknowledges the Mahidol University International Postdoctoral Fellowship. M.C.P. is also partially supported by Fundamental Fund: fiscal year 2024 by National Science Research and Innovation Fund (NSRF). M.C.P. also acknowledges Shinji Tsujikawa for fruitful discussions.
K. K. is supported by STFC grant ST/W001225/1. 
For the purpose of open access, we have applied a Creative Commons Attribution (CC BY) licence to any
Author Accepted Manuscript version arising. Supporting research data are available on request from the corresponding author.

\end{acknowledgments}
    
\appendix

\section{Equations for fluid}\label{appendix:fluid_action}
The dark matter action is described by
\begin{equation}
    S_{\rm fluid} = -\int {\rm d}^4 x \sqrt{-g} \left[ \rho_{c} (- n_{\mu } n^{\mu }) + n^{\mu } \nabla_{\mu }K \right]\,,
\end{equation}
where $K$ is a Lagrangian multiplier imposing the condition 
\begin{equation}
    \nabla_{\mu}n^{\mu} = 0\,.
\end{equation}

The equation of motion for the field $n^{\mu}$ is given by 
\begin{equation}
    \nabla_{\mu }K - 2 n_{\mu } \frac{\partial \rho_{c} \bigl(\tilde{n}\bigr)}{\partial \tilde{n}}.
\end{equation}
The variation with respect to $g^{\alpha \beta}$ gives the energy-momentum tensor 
\begin{equation}
   T_{\alpha \beta} \equiv \frac{2}{\sqrt{-g}} \frac{\delta S_{\rm fluid}}{\delta g^{\alpha \beta} } = g_{\alpha \beta } \rho_{c}  - 2 n_{\alpha } n_{\beta } \frac{\partial \rho_{c} }{\partial \tilde{n}} + 2 g_{\alpha \beta } n_{\gamma } n^{\gamma } \frac{\partial \rho_{c} }{\partial \tilde{n}}\,.
\end{equation}
The above expression for $T_{\alpha \beta}$ agrees with $T_{\alpha \beta}$ derived from the action without the constraint term. 

Assuming the homogeneous and isotropic background, we define
\begin{equation}
    K = K_{0}(t) + \delta K (t,x) \,,
\end{equation}
and $n^{\alpha}$ is defined by (\ref{nu_def}). Now we expand the action up to linear order to find the background equations of motion. The background equation of motion for $\delta K$ sets
\begin{equation}
    n_{0} = \frac{N}{a^3}\,,
\end{equation}
and equation of motion for $\delta n_{0}$ gives
\begin{equation}
    K_{0} = -2 \int a(\eta) n_{0}(\eta) \frac{\partial \rho_{c}}{\partial \tilde{n}} {\rm d} \eta \,. 
\end{equation}

To get the equations of motion for perturbation, we expand the action up to the second order. We make the following field redefinition
\begin{eqnarray}
    \delta n_{0} &=& -\alpha n_{0} + \frac{\rho_{c}}{2 n_{0} \frac{\partial \rho_{c}}{\partial \tilde{n}}} \delta \,.
\end{eqnarray}
The equation of motion for $n_{s}$ is given by 
\begin{equation}
    \delta K = \frac{2 N}{a^3} \frac{\partial \rho_{c}}{\partial \tilde{n}} \chi - \frac{2N}{a^2}  \frac{\partial \rho_{c}}{\partial \tilde{n}} \frac{\theta}{k^2}\,,
\end{equation}
where we used a field redefinition 
\begin{equation}
    n_{s} = -\frac{N }{a^3}\frac{a}{k^2}\theta \,.
\end{equation}

The remaining fields are $\delta$ and $\theta$. The equation of motion for $\delta$ is given by 
\begin{equation}
    E_{\delta} \equiv \dot{\theta} + aH(1 - 3 c_{s}^{2})\theta - \frac{c_{s}^{2} k^2 \rho}{\rho + P} \delta  + \frac{k^2}{a}\dot{\chi} +3 H k^2 c_{s}^{2} \chi -k^2 \alpha = 0 \,.
\end{equation}
Finally the equation of motion for the field $\theta$ is obtained as 
\begin{equation}
    E_{\theta} \equiv \dot{\delta} + 3aH \left(c_{s}^{2} - \frac{P}{\rho} \right) \delta + \left(1+ \frac{P}{\rho} \right) \theta - k^2 \left(1 + \frac{P}{\rho}  \right) \partial_{t} \left( E/a^2\right) + 3 \left(1 + \frac{P}{\rho} \right) \dot{\zeta} = 0 \,.
\end{equation}
Here we used the following definitions in the above expressions
\begin{eqnarray}
    c_{s}^2 &\equiv & \frac{\partial P}{\partial \rho} = \frac{\dot{P}}{\dot{\rho}} = 1 + \frac{2N^2 \rho_{c}^{\prime \prime }}{a^6 \rho_{c}^{\prime}} \,.
\end{eqnarray}

\section{Expression for $G_{\rm cc}$, $\Gamma_{\rm cc}$, and $\Gamma_{\rm cb}$}
Here we give the full lengthy expression for $G_{\rm cc}$, $\Gamma_{\rm cc}$, and $\Gamma_{\rm cb}$

\begin{eqnarray}
    G_{\rm cc} \equiv \frac{\Delta_{1}}{\Delta_{2}} G \,, \qquad \Gamma_{\rm cc} \equiv \frac{\Xi_{1}}{\Xi_{2}}\,, \qquad \Gamma_{\rm cb} \equiv \frac{\Upsilon_{1}}{\Upsilon_{2}} \,.
\end{eqnarray}
Since the expression for $\Xi_{1}$, and $\Xi_{2}$ is very large, the blow expression is with the limit $f_{\mathcal{Z}} \rightarrow 0$, $f_{\mathcal{ZZ}} \rightarrow 0$, $f_{\mathcal{YZ}} \rightarrow 0$ , $f_{\mathcal{ZE}} \rightarrow 0$.
\begin{eqnarray}
    \Delta_{1} &\equiv & 2 M_{\rm P}^2 a  G_{3, \mathcal{Y}}^4  \dot{\phi}  f_{,\mathcal{YE}} \phi  ^{11}+2 M_{\rm P}^2 a  G_{3, \mathcal{Y}}^2  \dot{\phi}  G_{, \mathcal{YY}} f_{, \mathcal{E}}^2 \phi  ^{10}+4 M_{\rm P}^2 a ^2 H  G_{3, \mathcal{Y}}^4 f_{, \mathcal{E}} \phi  ^{10}+2 M_{\rm P}^2 a  G_{3, \mathcal{Y}}^4  \dot{\phi}  f_{,\mathcal{ZE}} \phi  ^{10} \nonumber \\
    & & -4 M_{\rm P}^2 a  G_{3, \mathcal{Y}}^3  \dot{\phi}  f_{, \mathcal{E}} f_{,\mathcal{YE}} \phi  ^{10}-4 M_{\rm P}^2 a ^2 H  G_{3, \mathcal{Y}}^3 f_{, \mathcal{E}}^2 \phi  ^9-8 M_{\rm P}^4 G_{3, \mathcal{Y}}  \dot{\phi} ^2G_{, \mathcal{YY}}^2 f_{, \mathcal{E}} \phi  ^9+4 M_{\rm P}^4 G_{3, \mathcal{Y}}^2  \dot{\phi} ^2 G_{3,\mathcal{YYY}} f_{, \mathcal{E}} \phi  ^9 \nonumber \\
    & & -2 M_{\rm P}^2 a  G_{3, \mathcal{Y}}^2  \dot{\phi}  G_{, \mathcal{YY}} f_{, \mathcal{Z}} \phi  ^9-4 M_{\rm P}^2 a  G_{3, \mathcal{Y}}^3  \dot{\phi}  f_{, \mathcal{E}} f_{, \mathcal{ZE}} \phi  ^9 -4 M_{\rm P}^4 a  G_{3, \mathcal{Y}}^4 \dot{\phi}  f_{,\mathcal{YE}} \phi  ^9-4 M_{\rm P}^4 G_{3, \mathcal{Y}}^2  \dot{\phi} ^2   G_{, \mathcal{YY}} f_{,\mathcal{YE}} \phi  ^9 \nonumber \\
    & & +2 M_{\rm P}^2 a G_{3, \mathcal{Y}}^3  \dot{\phi}  f_{, \mathcal{YZ}} \phi  ^9-6 M_{\rm P}^4 a ^2 H ^2 \Omega_{c}  G_{3, \mathcal{Y}}^4 \phi  ^8+4 M_{\rm P}^4  \dot{\phi} ^2 G_{, \mathcal{YY}}^2   f_{, \mathcal{E}}^2 \phi  ^8+2 M_{\rm P}^2 a  G_{3, \mathcal{Y}}^3  \dot{\phi}    f_{, \mathcal{E}}^2 \phi  ^8+4 M_{\rm P}^4 a  G_{3, \mathcal{Y}}^2  \dot{\phi}   G_{, \mathcal{YY}} f_{, \mathcal{E}}^2 \phi^8 \nonumber \\
    & & -4 M_{\rm P}^4 G_{3, \mathcal{Y}}  \dot{\phi} ^2 G_{3,\mathcal{YYY}} f_{, \mathcal{E}}^2 \phi  ^8-6 M_{\rm P}^4 a  H ^2 \Omega_{c}  G_{3, \mathcal{Y}}^2  \dot{\phi}  G_{, \mathcal{YY}} \phi^8-8 M_{\rm P}^4 a ^2 H  G_{3, \mathcal{Y}}^4 f_{, \mathcal{E}} \phi  ^8-16 M_{\rm P}^4 a  H  G_{3, \mathcal{Y}}^2  \dot{\phi}  G_{, \mathcal{YY}} f_{, \mathcal{E}} \phi^8 \nonumber \\
    & & +4 M_{\rm P}^4 G_{3, \mathcal{Y}}^2 G_{, \mathcal{YY}}  \dot{\phi}' f_{, \mathcal{E}} \phi  ^8+4 M_{\rm P}^2 a ^2 H  G_{3, \mathcal{Y}}^3 f_{,\mathcal{Z}} \phi  ^8-4 M_{\rm P}^4 a  G_{3, \mathcal{Y}}^4  \dot{\phi} f_{, \mathcal{ZE}} \phi  ^8-4 M_{\rm P}^4 G_{3, \mathcal{Y}}^2  \dot{\phi} ^2 G_{, \mathcal{YY}} f_{, \mathcal{ZE}} \phi  ^8+2 M_{\rm P}^2 a G_{3, \mathcal{Y}}^3  \dot{\phi}  f_{, \mathcal{ZZ}}\phi^8 \nonumber \\
    & & -4 M_{\rm P}^4 a  H G_{3, \mathcal{Y}}^3  \dot{\phi}  f_{,\mathcal{YE}} \phi  ^8+8 M_{\rm P}^4 a G_{3, \mathcal{Y}}^3  \dot{\phi}  f_{, \mathcal{E}} f_{, \mathcal{YE}} \phi  ^8+8 M_{\rm P}^4 G_{3, \mathcal{Y}}  \dot{\phi} ^2 G_{, \mathcal{YY}} f_{, \mathcal{E}} f_{,\mathcal{YE}} \phi  ^8-6 M_{\rm P}^4 a ^2 H ^3\Omega_{c}  G_{3, \mathcal{Y}}^3 \phi  ^7 \nonumber \\
    & & -8 M_{\rm P}^4 a  G_{3, \mathcal{Y}}  \dot{\phi} G_{, \mathcal{YY}} f_{, \mathcal{E}}^3 \phi  ^7+24 M_{\rm P}^4 a ^2 H G_{3, \mathcal{Y}}^3 f_{, \mathcal{E}}^2 \phi  ^7+12 M_{\rm P}^4 a  H G_{3, \mathcal{Y}}  \dot{\phi}  G_{, \mathcal{YY}} f_{, \mathcal{E}}^2 \phi  ^7-4 M_{\rm P}^4 G_{3, \mathcal{Y}} G_{, \mathcal{YY}}  \ddot{\phi} f_{, \mathcal{E}}^2 \phi^7 \nonumber \\
    & & -8 M_{\rm P}^4 a ^2 H ^2 G_{3, \mathcal{Y}}^3 f_{, \mathcal{E}} \phi  ^7+12 M_{\rm P}^4 a ^2 H ^2 \Omega_{c}  G_{3, \mathcal{Y}}^3 f_{, \mathcal{E}} \phi  ^7+4 M_{\rm P}^4 a  G_{3, \mathcal{Y}}^3 \dot{H} f_{, \mathcal{E}} \phi  ^7-8 M_{\rm P}^4 a G_{3, \mathcal{Y}}^4  \dot{\phi}    f_{, \mathcal{E}} \phi  ^7-16 M_{\rm P}^4 G_{3, \mathcal{Y}}^2  \dot{\phi} ^2 G_{, \mathcal{YY}} f_{, \mathcal{E}} \phi  ^7 \nonumber \\
    & & -4 M_{\rm P}^4  \dot{\phi} ^2 G_{, \mathcal{YY}}^2 f_{,\mathcal{Z}} \phi  ^7-4 M_{\rm P}^2 a G_{3, \mathcal{Y}}^3  \dot{\phi}  f_{,\mathcal{Z}} \phi  ^7+4 M_{\rm P}^4 a G_{3, \mathcal{Y}}^2  \dot{\phi}  G_{, \mathcal{YY}} f_{, \mathcal{Z}} \phi  ^7+4 M_{\rm P}^4 G_{3, \mathcal{Y}}  \dot{\phi} ^2 G_{3,\mathcal{YYY}} f_{,\mathcal{Z}} \phi  ^7 \nonumber \\
    & & -4 M_{\rm P}^4 a  H  G_{3, \mathcal{Y}}^3  \dot{\phi} f_{, \mathcal{ZE}} \phi  ^7+8 M_{\rm P}^4 a  G_{3, \mathcal{Y}}^3  \dot{\phi} f_{, \mathcal{E}} f_{, \mathcal{ZE}} \phi  ^7+8 M_{\rm P}^4 G_{3, \mathcal{Y}}  \dot{\phi} ^2 G_{, \mathcal{YY}} f_{, \mathcal{E}} f_{, \mathcal{ZE}} \phi  ^7-8 M_{\rm P}^4 G_{3, \mathcal{Y}}^3  \dot{\phi} ^2 f_{,\mathcal{YE}} \phi  ^7 \nonumber \\
    & & -4 M_{\rm P}^4 a  G_{3, \mathcal{Y}}^2  \dot{\phi} f_{, \mathcal{E}}^2 f_{,\mathcal{YE}} \phi  ^7+8 M_{\rm P}^4 a  H G_{3, \mathcal{Y}}^2  \dot{\phi}  f_{, \mathcal{E}} f_{, \mathcal{YE}} \phi  ^7-4 M_{\rm P}^4 a  G_{3, \mathcal{Y}}^3  \dot{\phi}  f_{, \mathcal{YZ}} \phi  ^7-4 M_{\rm P}^4 G_{3, \mathcal{Y}}  \dot{\phi} ^2 G_{, \mathcal{YY}} f_{, \mathcal{YZ}} \phi  ^7 \nonumber \\
    & & +12 M_{\rm P}^6 a ^2 H ^2 \Omega_{c}  G_{3, \mathcal{Y}}^4 \phi  ^6+4 M_{\rm P}^4 a   \dot{\phi}  G_{, \mathcal{YY}} f_{, \mathcal{E}}^4 \phi  ^6-24 M_{\rm P}^4 a ^2 H  G_{3, \mathcal{Y}}^2 f_{, \mathcal{E}}^3 \phi  ^6+12 M_{\rm P}^6 H ^2 \Omega_{c}   \dot{\phi} ^2 G_{, \mathcal{YY}}^2 \phi  ^6+8 M_{\rm P}^4a ^2 H ^2 G_{3, \mathcal{Y}}^2 f_{, \mathcal{E}}^2 \phi  ^6 \nonumber \\
    & & -6 M_{\rm P}^4 a ^2 H ^2 \Omega_{c}  G_{3, \mathcal{Y}}^2 f_{, \mathcal{E}}^2 \phi  ^6-4 M_{\rm P}^4 a  G_{3, \mathcal{Y}}^2 \dot{H}  f_{, \mathcal{E}}^2 \phi  ^6+28 M_{\rm P}^4 a G_{3, \mathcal{Y}}^3  \dot{\phi}  f_{, \mathcal{E}}^2 \phi  ^6+4 M_{\rm P}^4 G_{3, \mathcal{Y}}  \dot{\phi} ^2 G_{, \mathcal{YY}} f_{, \mathcal{E}}^2 \phi  ^6 \nonumber \\
    & & -12 M_{\rm P}^4 a  H ^2 \Omega_{c}  G_{3, \mathcal{Y}}^3  \dot{\phi}  \phi  ^6+24 M_{\rm P}^6 a  H ^2 \Omega_{c}  G_{3, \mathcal{Y}}^2  \dot{\phi}  G_{, \mathcal{YY}} \phi  ^6-36 M_{\rm P}^4 a  H  G_{3, \mathcal{Y}}^3  \dot{\phi}  f_{, \mathcal{E}} \phi  ^6+8 M_{\rm P}^4 G_{3, \mathcal{Y}}^3  \dot{\phi}'  f_{, \mathcal{E}} \phi  ^6 \nonumber \\
    & & -8 M_{\rm P}^4 a ^2 H  G_{3, \mathcal{Y}}^3 f_{,\mathcal{Z}} \phi  ^6-12 M_{\rm P}^4 a  H G_{3, \mathcal{Y}}  \dot{\phi}  G_{, \mathcal{YY}} f_{,\mathcal{Z}} \phi  ^6+4 M_{\rm P}^4 G_{3, \mathcal{Y}} G_{, \mathcal{YY}}  \dot{\phi}'  f_{, \mathcal{Z}} \phi  ^6+8 M_{\rm P}^4 a  G_{3, \mathcal{Y}}  \dot{\phi}  G_{, \mathcal{YY}}f_{, \mathcal{E}} f_{,\mathcal{Z}} \phi  ^6 \nonumber \\
    & & -8 M_{\rm P}^4 G_{3, \mathcal{Y}}^3  \dot{\phi} ^2 f_{, \mathcal{ZE}} \phi  ^6-4 M_{\rm P}^4 a G_{3, \mathcal{Y}}^2  \dot{\phi}  f_{, \mathcal{E}}^2f_{, \mathcal{ZE}} \phi  ^6+8 M_{\rm P}^4 a  H  G_{3, \mathcal{Y}}^2  \dot{\phi}  f_{, \mathcal{E}} f_{, \mathcal{ZE}} \phi  ^6-4 M_{\rm P}^4 a G_{3, \mathcal{Y}}^3  \dot{\phi}  f_{, \mathcal{ZZ}}\phi^6 \nonumber \\
    & & - 4 M_{\rm P}^4 G_{3, \mathcal{Y}} \dot{\phi}^2 G_{, \mathcal{YY}} f_{, \mathcal{ZZ}}\phi^6 + 16 M_{\rm P}^4 G_{3, \mathcal{Y}}^2  \dot{\phi}^2 f_{, \mathcal{E}} f_{, \mathcal{YE}} \phi^6 - 4 M_{\rm P}^4 a  G_{3, \mathcal{Y}}^2  \dot{\phi}  f_{, \mathcal{Z}} f_{,\mathcal{YE}} \phi^6 - 4 M_{\rm P}^4 a  H  G_{3, \mathcal{Y}}^2 \dot{\phi}  f_{, \mathcal{YZ}} \phi  ^6 \nonumber \\
    & & +4 M_{\rm P}^4 a  G_{3, \mathcal{Y}}^2  \dot{\phi} f_{, \mathcal{E}} f_{, \mathcal{YZ}} \phi  ^6+8 M_{\rm P}^4 a ^2 H G_{3, \mathcal{Y}} f_{, \mathcal{E}}^4 \phi  ^5+24 M_{\rm P}^6 a ^2 H ^3 \Omega_{c}  G_{3, \mathcal{Y}}^3 \phi  ^5-32 M_{\rm P}^4 a  G_{3, \mathcal{Y}}^2  \dot{\phi} f_{, \mathcal{E}}^3 \phi  ^5+32 M_{\rm P}^4 a  H  G_{3, \mathcal{Y}}^2  \dot{\phi} f_{, \mathcal{E}}^2 \phi^5 \nonumber \\
    & & -8 M_{\rm P}^4 G_{3, \mathcal{Y}}^2  \ddot{\phi} f_{, \mathcal{E}}^2 \phi  ^5+24 M_{\rm P}^6 a  H ^3 \Omega_{c}  G_{3, \mathcal{Y}}  \dot{\phi}  G_{, \mathcal{YY}} \phi  ^5-48 M_{\rm P}^6 a ^2 H ^2 \Omega_{c}  G_{3, \mathcal{Y}}^3 f_{, \mathcal{E}} \phi  ^5-32 M_{\rm P}^4 G_{3, \mathcal{Y}}^3 \dot{\phi} ^2 f_{, \mathcal{E}} \phi  ^5 \nonumber \\
    & & - 48 M_{\rm P}^6 a  H ^2 \Omega_{c}  G_{3, \mathcal{Y}}  \dot{\phi}  G_{, \mathcal{YY}} f_{, \mathcal{E}} \phi  ^5-8 M_{\rm P}^4 a ^2 H ^2 G_{3, \mathcal{Y}}^2 f_{,\mathcal{Z}} \phi  ^5+6 M_{\rm P}^4 a ^2 H ^2 \Omega_{c}  G_{3, \mathcal{Y}}^2 f_{,\mathcal{Z}} \phi  ^5-8 M_{\rm P}^4 a   \dot{\phi}  G_{, \mathcal{YY}} f_{, \mathcal{E}}^2 f_{, \mathcal{Z}} \phi  ^5 \nonumber \\
    & & +4 M_{\rm P}^4 a  G_{3, \mathcal{Y}}^2 \dot{H}  f_{, \mathcal{Z}} \phi  ^5+24 M_{\rm P}^4 a ^2 H  G_{3, \mathcal{Y}}^2 f_{, \mathcal{E}} f_{,\mathcal{Z}} \phi  ^5+16 M_{\rm P}^4 G_{3, \mathcal{Y}}^2  \dot{\phi} ^2 f_{, \mathcal{E}} f_{, \mathcal{ZE}} \phi  ^5-4 M_{\rm P}^4 a G_{3, \mathcal{Y}}^2  \dot{\phi}  f_{,\mathcal{Z}} f_{, \mathcal{ZE}} \phi  ^5 \nonumber \\
    & & -4 M_{\rm P}^4 a  H  G_{3, \mathcal{Y}}^2  \dot{\phi}  f_{, \mathcal{ZZ}} \phi  ^5+4 M_{\rm P}^4 a  G_{3, \mathcal{Y}}^2  \dot{\phi}  f_{, \mathcal{E}} f_{, \mathcal{ZZ}}\phi  ^5-8 M_{\rm P}^4 G_{3, \mathcal{Y}}^2  \dot{\phi} ^2 f_{, \mathcal{YZ}} \phi  ^5+12 M_{\rm P}^4 a  G_{3, \mathcal{Y}}  \dot{\phi} f_{, \mathcal{E}}^4 \phi  ^4+12 M_{\rm P}^6 a ^2 H ^4 \Omega_{c}  G_{3, \mathcal{Y}}^2 \phi^4 \nonumber \\
    & & +72 M_{\rm P}^6 a ^2 H ^2 \Omega_{c}  G_{3, \mathcal{Y}}^2 f_{, \mathcal{E}}^2 \phi  ^4+24 M_{\rm P}^4 G_{3, \mathcal{Y}}^2  \dot{\phi} ^2 f_{, \mathcal{E}}^2 \phi  ^4+24 M_{\rm P}^6 a  H ^2 \Omega_{c}   \dot{\phi}  G_{, \mathcal{YY}} f_{, \mathcal{E}}^2 \phi  ^4+4 M_{\rm P}^4 a   \dot{\phi}  G_{, \mathcal{YY}} f_{, \mathcal{Z}}^2 \phi  ^4\nonumber \\
    & & +48 M_{\rm P}^6 a  H ^2 \Omega_{c}  G_{3, \mathcal{Y}}^3  \dot{\phi}  \phi  ^4 \nonumber +48M_{\rm P}^6 H ^2 \Omega_{c}  G_{3, \mathcal{Y}}  \dot{\phi} ^2 G_{, \mathcal{YY}} \phi  ^4-48M_{\rm P}^6 a ^2 H ^3 \Omega_{c}  G_{3, \mathcal{Y}}^2 f_{, \mathcal{E}} \phi^4-16 M_{\rm P}^4 a ^2 H  G_{3, \mathcal{Y}} f_{, \mathcal{E}}^2 f_{,\mathcal{Z}} \phi  ^4\nonumber \\
    & & -28 M_{\rm P}^4 a  H  G_{3, \mathcal{Y}}^2  \dot{\phi} f_{,\mathcal{Z}} \phi  ^4+8 M_{\rm P}^4 G_{3, \mathcal{Y}}^2  \dot{\phi}' f_{,\mathcal{Z}} \phi  ^4+28 M_{\rm P}^4 a  G_{3, \mathcal{Y}}^2  \dot{\phi} f_{, \mathcal{E}} f_{,\mathcal{Z}} \phi  ^4-8 M_{\rm P}^4 G_{3, \mathcal{Y}}^2  \dot{\phi} ^2 f_{, \mathcal{ZZ}}\phi  ^4 \nonumber \\
    & & -48 M_{\rm P}^6 a ^2 H ^2 \Omega_{c}  G_{3, \mathcal{Y}} f_{, \mathcal{E}}^3 \phi  ^3+24 M_{\rm P}^6 a ^2 H ^3 \Omega_{c}  G_{3, \mathcal{Y}} f_{, \mathcal{E}}^2 \phi  ^3+8 M_{\rm P}^4 a ^2 H  G_{3, \mathcal{Y}} f_{,\mathcal{Z}}^2 \phi  ^3+48 M_{\rm P}^6 a  H ^3 \Omega_{c}  G_{3, \mathcal{Y}}^2  \dot{\phi}  \phi  ^3 \nonumber \\
    & & -96 M_{\rm P}^6 a  H ^2 \Omega_{c} G_{3, \mathcal{Y}}^2  \dot{\phi}  f_{, \mathcal{E}} \phi  ^3-24 M_{\rm P}^6 a ^2 H ^2 \Omega_{c}  G_{3, \mathcal{Y}}^2 f_{,\mathcal{Z}} \phi  ^3-16 M_{\rm P}^4 G_{3, \mathcal{Y}}^2  \dot{\phi} ^2 f_{,\mathcal{Z}} \phi  ^3 \nonumber \\
    & & -24 M_{\rm P}^4 a G_{3, \mathcal{Y}}  \dot{\phi}  f_{, \mathcal{E}}^2 f_{, \mathcal{Z}} \phi  ^3-24 M_{\rm P}^6 a  H ^2 \Omega_{c}   \dot{\phi}  G_{, \mathcal{YY}} f_{,\mathcal{Z}} \phi  ^3+12 M_{\rm P}^6 a ^2 H ^2 \Omega_{c}  f_{, \mathcal{E}}^4 \phi^2+48 M_{\rm P}^6 H ^2 \Omega_{c}  G_{3, \mathcal{Y}}^2  \dot{\phi} ^2 \phi  ^2\nonumber \\
    & & +48 M_{\rm P}^6 a  H ^2 \Omega_{c}  G_{3, \mathcal{Y}}  \dot{\phi}  f_{, \mathcal{E}}^2 \phi  ^2+12 M_{\rm P}^4 a  G_{3, \mathcal{Y}}  \dot{\phi}  f_{,\mathcal{Z}}^2 \phi  ^2-24 M_{\rm P}^6 a ^2 H ^3 \Omega_{c}  G_{3, \mathcal{Y}} f_{,\mathcal{Z}} \phi  ^2+48 M_{\rm P}^6 a ^2 H ^2 \Omega_{c}  G_{3, \mathcal{Y}} f_{, \mathcal{E}} f_{,\mathcal{Z}} \phi  ^2 \nonumber \\
    & & -24 M_{\rm P}^6 a ^2 H ^2 \Omega_{c}  f_{, \mathcal{E}}^2 f_{,\mathcal{Z}} \phi  -48 M_{\rm P}^6 a  H ^2 \Omega_{c} G_{3, \mathcal{Y}}  \dot{\phi}  f_{,\mathcal{Z}} \phi  +12 M_{\rm P}^6 a ^2 H ^2 \Omega_{c}  f_{,\mathcal{Z}}^2 \,,
\end{eqnarray}
\begin{eqnarray}
    \Delta_{2} & \equiv & -a ^2 G_{3,\mathcal{Y}}^4 f_{,\mathcal{E}}^2 \phi  ^{12}+a ^2   G_{3,\mathcal{Y}}^4 f_{3,\mathcal{Z}} \phi  ^{11}+3 M_{\rm P}^2 a ^2 H ^2  \Omega_{c}  G_{3,\mathcal{Y}}^4 \phi  ^{10}+2 M_{\rm P}^2 a ^2 G_{3,\mathcal{Y}}^4 f_{,\mathcal{E}}^2 \phi  ^{10}+4 M_{\rm P}^2 a  G_{3,\mathcal{Y}}^2 \dot{\phi} G_{3,\mathcal{YY}} f_{,\mathcal{E}}^2 \phi  ^{10} \nonumber \\
    & & -4 M_{\rm P}^2 a ^2 G_{3,\mathcal{Y}}^3 f_{,\mathcal{E}}^3 \phi  ^9+4 M_{\rm P}^2 a ^2 H G_{3,\mathcal{Y}}^3 f_{,\mathcal{E}}^2 \phi  ^9-4 M_{\rm P}^2 a ^2 G_{3,\mathcal{Y}}^4 f_{3,\mathcal{Z}} \phi  ^9-4 M_{\rm P}^2 a G_{3,\mathcal{Y}}^2 \dot{\phi}  G_{3,\mathcal{YY}} f_{,\mathcal{Z}} \phi^9 \nonumber \\
    & & -12 M_{\rm P}^4 a ^2 H ^2 \Omega_{c}  G_{3,\mathcal{Y}}^4 \phi  ^8+2 M_{\rm P}^2 a ^2 G_{3,\mathcal{Y}}^2 f_{,\mathcal{E}}^4 \phi  ^8-4 M_{\rm P}^4 \dot{\phi} ^2 G_{3,\mathcal{YY}}^2 f_{,\mathcal{E}}^2 \phi  ^8+8 M_{\rm P}^2 a G_{3,\mathcal{Y}}^3 \dot{\phi}  f_{,\mathcal{E}}^2 \phi  ^8 \nonumber \\
    & & -4 M_{\rm P}^4 a G_{3,\mathcal{Y}}^2 \dot{\phi}  G_{3,\mathcal{YY}} f_{,\mathcal{E}}^2 \phi  ^8-12 M_{\rm P}^4 a  H ^2 \Omega_{c}  G_{3,\mathcal{Y}}^2 \dot{\phi}  G_{3,\mathcal{YY}} \phi  ^8-4 M_{\rm P}^2 a ^2 H  G_{3,\mathcal{Y}}^3 f_{,\mathcal{Z}} \phi  ^8+4 M_{\rm P}^2 a ^2 G_{3,\mathcal{Y}}^3 f_{,\mathcal{E}}
   f_{3,\mathcal{Z}} \phi  ^8 \nonumber \\
   & & -12 M_{\rm P}^4 a ^2 H ^3 \Omega_{c}  G_{3,\mathcal{Y}}^3 \phi  ^7+8 M_{\rm P}^4 a  G_{3,\mathcal{Y}} \dot{\phi}  G_{3,\mathcal{YY}} f_{,\mathcal{E}}^3 \phi  ^7-4 M_{\rm P}^4 a ^2 H  G_{3,\mathcal{Y}}^3 f_{,\mathcal{E}}^2 \phi  ^7-8 M_{\rm P}^4 a  H  G_{3,\mathcal{Y}} \dot{\phi} G_{3,\mathcal{YY}} f_{,\mathcal{E}}^2 \phi  ^7\nonumber \\
   & & +24 M_{\rm P}^4 a ^2 H ^2 \Omega_{c}  G_{3,\mathcal{Y}}^3 f_{,\mathcal{E}} \phi  ^7+4 M_{\rm P}^4 a ^2 G_{3,\mathcal{Y}}^4 f_{3,\mathcal{Z}} \phi  ^7+4 M_{\rm P}^4 \dot{\phi} ^2 G_{3,\mathcal{YY}}^2 f_{3,\mathcal{Z}} \phi  ^7-4 M_{\rm P}^2 a ^2 G_{3,\mathcal{Y}}^2 f_{,\mathcal{E}}^2 f_{,\mathcal{Z}} \phi  ^7\nonumber \\
   & & -8 M_{\rm P}^2 a  G_{3,\mathcal{Y}}^3 \dot{\phi}  f_{,\mathcal{Z}} \phi  ^7+8 M_{\rm P}^4 a  G_{3,\mathcal{Y}}^2 \dot{\phi}  G_{3,\mathcal{YY}} f_{3,\mathcal{Z}} \phi  ^7+12 M_{\rm P}^6 a ^2 H ^2 \Omega_{c}  G_{3,\mathcal{Y}}^4 \phi  ^6-4 M_{\rm P}^4 a  \dot{\phi}  G_{3,\mathcal{YY}} f_{,\mathcal{E}}^4 \phi  ^6 \nonumber \\
   & & +8 M_{\rm P}^4 a ^2 H  G_{3,\mathcal{Y}}^2 f_{,\mathcal{E}}^3 \phi  ^6+12 M_{\rm P}^6 H ^2 \Omega_{c}  \dot{\phi} ^2 G_{3,\mathcal{YY}}^2 \phi  ^6-4 M_{\rm P}^4 a ^2 H ^2 G_{3,\mathcal{Y}}^2 f_{,\mathcal{E}}^2 \phi  ^6-12 M_{\rm P}^4 a ^2 H ^2 \Omega_{c}  G_{3,\mathcal{Y}}^2 f_{,\mathcal{E}}^2 \phi  ^6 \nonumber \\
   & & -8 M_{\rm P}^4 a  G_{3,\mathcal{Y}}^3 \dot{\phi} f_{,\mathcal{E}}^2 \phi  ^6-16 M_{\rm P}^4 G_{3,\mathcal{Y}} \dot{\phi} ^2 G_{3,\mathcal{YY}} f_{,\mathcal{E}}^2 \phi  ^6+2 M_{\rm P}^2 a ^2 G_{3,\mathcal{Y}}^2 f_{3,\mathcal{Z}}^2 \phi  ^6-24 M_{\rm P}^4 a  H ^2 \Omega_{c}  G_{3,\mathcal{Y}}^3 \dot{\phi}  \phi  ^6 \nonumber \\
   & & +24 M_{\rm P}^6 a  H ^2 \Omega_{c}  G_{3,\mathcal{Y}}^2 \dot{\phi}G_{3,\mathcal{YY}} \phi  ^6+8 M_{\rm P}^4 a ^2 H  G_{3,\mathcal{Y}}^3 f_{3,\mathcal{Z}} \phi  ^6+8 M_{\rm P}^4 a  H G_{3,\mathcal{Y}} \dot{\phi}  G_{3,\mathcal{YY}} f_{3,\mathcal{Z}} \phi  ^6-8 M_{\rm P}^4 a ^2 G_{3,\mathcal{Y}}^3 f_{,\mathcal{E}} f_{3,\mathcal{Z}} \phi  ^6 \nonumber \\
   & & -8 M_{\rm P}^4 a  G_{3,\mathcal{Y}} \dot{\phi} G_{3,\mathcal{YY}} f_{,\mathcal{E}} f_{,\mathcal{Z}} \phi  ^6-4 M_{\rm P}^4 a ^2 H  G_{3,\mathcal{Y}} f_{,\mathcal{E}}^4\phi  ^5+24 M_{\rm P}^6 a ^2 H ^3 \Omega_{c}  G_{3,\mathcal{Y}}^3 \phi  ^5+16 M_{\rm P}^4 a G_{3,\mathcal{Y}}^2 \dot{\phi}  f_{,\mathcal{E}}^3 \phi  ^5 \nonumber \\
   & & -16 M_{\rm P}^4 a  H G_{3,\mathcal{Y}}^2 \dot{\phi}  f_{,\mathcal{E}}^2 \phi  ^5+24 M_{\rm P}^6 a  H ^3\Omega_{c}  G_{3,\mathcal{Y}} \dot{\phi}  G_{3,\mathcal{YY}} \phi  ^5-48 M_{\rm P}^6 a ^2 H ^2 \Omega_{c}  G_{3,\mathcal{Y}}^3 f_{,\mathcal{E}} \phi  ^5 \nonumber \\
   & & -48 M_{\rm P}^6 a  H ^2 \Omega_{c}  G_{3,\mathcal{Y}} \dot{\phi} G_{3,\mathcal{YY}} f_{,\mathcal{E}} \phi  ^5+4 M_{\rm P}^4 a ^2 H ^2 G_{3,\mathcal{Y}}^2 f_{3,\mathcal{Z}} \phi  ^5+12 M_{\rm P}^4 a ^2 H ^2 \Omega_{c}  G_{3,\mathcal{Y}}^2 f_{3,\mathcal{Z}} \phi  ^5+4 M_{\rm P}^4 a ^2 G_{3,\mathcal{Y}}^2 f_{,\mathcal{E}}^2 f_{3,\mathcal{Z}} \phi  ^5 \nonumber \\
   & & +8 M_{\rm P}^4 a  \dot{\phi}  G_{3,\mathcal{YY}} f_{,\mathcal{E}}^2 f_{,\mathcal{Z}} \phi  ^5+16 M_{\rm P}^4 a  G_{3,\mathcal{Y}}^3 \dot{\phi}  f_{,\mathcal{Z}} \phi  ^5+16 M_{\rm P}^4 G_{3,\mathcal{Y}} \dot{\phi} ^2 G_{3,\mathcal{YY}} f_{3,\mathcal{Z}} \phi  ^5-8 M_{\rm P}^4 a ^2 H  G_{3,\mathcal{Y}}^2 f_{,\mathcal{E}} f_{3,\mathcal{Z}} \phi  ^5 \nonumber \\
   & & -8 M_{\rm P}^4 a G_{3,\mathcal{Y}} \dot{\phi}  f_{,\mathcal{E}}^4 \phi  ^4+12 M_{\rm P}^6 a ^2 H ^4 \Omega_{c}  G_{3,\mathcal{Y}}^2 \phi  ^4+72 M_{\rm P}^6 a ^2 H ^2 \Omega_{c}  G_{3,\mathcal{Y}}^2 f_{,\mathcal{E}}^2 \phi  ^4-16 M_{\rm P}^4 G_{3,\mathcal{Y}}^2 \dot{\phi} ^2 f_{,\mathcal{E}}^2 \phi  ^4 \nonumber \\
   & & +24 M_{\rm P}^6 a  H ^2 \Omega_{c}  \dot{\phi} G_{3,\mathcal{YY}} f_{,\mathcal{E}}^2 \phi  ^4-4 M_{\rm P}^4 a ^2 G_{3,\mathcal{Y}}^2 f_{3,\mathcal{Z}}^2 \phi  ^4-4 M_{\rm P}^4 a  \dot{\phi} G_{3,\mathcal{YY}} f_{3,\mathcal{Z}}^2 \phi  ^4+48 M_{\rm P}^6 a  H ^2 \Omega_{c}  G_{3,\mathcal{Y}}^3 \dot{\phi}  \phi  ^4 \nonumber \\
   & & +48 M_{\rm P}^6 H ^2 \Omega_{c}  G_{3,\mathcal{Y}} \dot{\phi} ^2 G_{3,\mathcal{YY}} \phi  ^4-48 M_{\rm P}^6 a ^2 H ^3 \Omega_{c}  G_{3,\mathcal{Y}}^2 f_{,\mathcal{E}} \phi  ^4+8 M_{\rm P}^4 a ^2 H  G_{3,\mathcal{Y}} f_{,\mathcal{E}}^2 f_{3,\mathcal{Z}} \phi  ^4+16 M_{\rm P}^4 a  H  G_{3,\mathcal{Y}}^2 \dot{\phi}  f_{3,\mathcal{Z}} \phi  ^4 \nonumber \\
   & & -16 M_{\rm P}^4 a  G_{3,\mathcal{Y}}^2 \dot{\phi}  f_{,\mathcal{E}} f_{,\mathcal{Z}} \phi  ^4-48 M_{\rm P}^6 a ^2 H ^2 \Omega_{c}  G_{3,\mathcal{Y}} f_{,\mathcal{E}}^3 \phi  ^3+24 M_{\rm P}^6 a ^2 H ^3 \Omega_{c}  G_{3,\mathcal{Y}} f_{,\mathcal{E}}^2 \phi  ^3-4 M_{\rm P}^4 a ^2 H  G_{3,\mathcal{Y}} f_{3,\mathcal{Z}}^2 \phi  ^3 \nonumber \\
   & & +48 M_{\rm P}^6 a  H ^3 \Omega_{c}  G_{3,\mathcal{Y}}^2 \dot{\phi}  \phi  ^3-96 M_{\rm P}^6 a  H ^2 \Omega_{c}  G_{3,\mathcal{Y}}^2 \dot{\phi} f_{,\mathcal{E}} \phi  ^3-24 M_{\rm P}^6 a ^2 H ^2 \Omega_{c}  G_{3,\mathcal{Y}}^2 f_{3,\mathcal{Z}} \phi  ^3+16 M_{\rm P}^4 G_{3,\mathcal{Y}}^2 \dot{\phi} ^2 f_{3,\mathcal{Z}} \phi  ^3 \nonumber \\
   & & +16 M_{\rm P}^4 a  G_{3,\mathcal{Y}} \dot{\phi}   f_{,\mathcal{E}}^2 f_{3,\mathcal{Z}} \phi  ^3-24 M_{\rm P}^6 a  H ^2 \Omega_{c}  \dot{\phi}  G_{3,\mathcal{YY}} f_{3,\mathcal{Z}} \phi  ^3+12 M_{\rm P}^6 a ^2 H ^2 \Omega_{c}  f_{,\mathcal{E}}^4 \phi  ^2 +48 M_{\rm P}^6 H ^2 \Omega_{c} G_{3,\mathcal{Y}}^2 \dot{\phi} ^2 \phi  ^2 \nonumber \\
   & & +48 M_{\rm P}^6 a  H ^2 \Omega_{c}  G_{3,\mathcal{Y}} \dot{\phi}  f_{,\mathcal{E}}^2 \phi  ^2-8 M_{\rm P}^4 a  G_{3,\mathcal{Y}} \dot{\phi}  f_{3,\mathcal{Z}}^2 \phi  ^2-24 M_{\rm P}^6 a ^2 H ^3 \Omega_{c} G_{3,\mathcal{Y}} f_{,\mathcal{Z}} \phi  ^2+48 M_{\rm P}^6 a ^2 H ^2 \Omega_{c}  G_{3,\mathcal{Y}} f_{,\mathcal{E}} f_{,\mathcal{Z}} \phi  ^2 \nonumber \\
   & & -24 M_{\rm P}^6 a ^2 H ^2 \Omega_{c}  f_{,\mathcal{E}}^2f_{3,\mathcal{Z}} \phi  -48 M_{\rm P}^6 a  H ^2 \Omega_{c}  G_{3,\mathcal{Y}} \dot{\phi}  f_{3,\mathcal{Z}} \phi  +12 M_{\rm P}^6 a ^2 H ^2 \Omega_{c}  f_{,\mathcal{Z}})^2 \,,
\end{eqnarray}

\begin{eqnarray}
    \Upsilon_{1} & \equiv & 3 M_{\rm P}^2 a H^2 \Omega_{b} \phi ^5 f_{,\mathcal{E}} G_{3,\mathcal{Y}}^2-3 M_{\rm P}^2 a H^2 \Omega_{b} \phi ^4 f_{,\mathcal{E}}^2 G_{3,\mathcal{Y}}+3 M_{\rm P}^2 a H^2 \Omega_{b} \phi ^3 f_{,\mathcal{Z}}
   G_{3,\mathcal{Y}} \,,
\end{eqnarray}

\begin{eqnarray}
    \Upsilon_{2} & \equiv & -12 M_{\rm P}^4 a  H ^2 \Omega_{c}  \phi  ^2 f_{,\mathcal{E}} G_{3,\mathcal{Y}}-2 M_{\rm P}^2 a  H  \phi  ^4 f_{,\mathcal{E}}^2G_{3,\mathcal{Y}}+2 M_{\rm P}^2 a  H  \phi  ^3 f_{,\mathcal{Z}}G_{3,\mathcal{Y}}+2 M_{\rm P}^2 a \phi  ^4 f_{,\mathcal{Z}}G_{3,\mathcal{Y}}^2+a  \phi  ^7 f_{,\mathcal{E}}^2G_{3,\mathcal{Y}}^2-a  \phi  ^6 f_{,\mathcal{Z}}G_{3,\mathcal{Y}}^2 \nonumber \\
    & & +6 M_{\rm P}^4 a  H ^2 \Omega_{c}  \phi   f_{,\mathcal{E}}^2-6 M_{\rm P}^4 a  H ^2 \Omega_{c}  f_{,\mathcal{Z}}-2 M_{\rm P}^2 \phi^5 \dot{\phi}  f_{,\mathcal{E}}^2 G_{3,\mathcal{YY}}+2 M_{\rm P}^2 \phi  ^4 \dot{\phi}  f_{,\mathcal{Z}} G_{3,\mathcal{YY}}-4 M_{\rm P}^2 \phi  ^3 \dot{\phi}    f_{,\mathcal{E}}^2G_{3,\mathcal{Y}}+4 M_{\rm P}^2 \phi  ^2 \dot{\phi}    f_{,\mathcal{Z}}G_{3,\mathcal{Y}}\nonumber \\
    & & +6 M_{\rm P}^4 a  H ^2 \Omega_{c}  \phi^3G_{3,\mathcal{Y}}^2+6 M_{\rm P}^4 a  H ^3 \Omega_{c}  \phi  ^2G_{3,\mathcal{Y}}-3 M_{\rm P}^2 a  H ^2 \Omega_{c}  \phi  ^5G_{3,\mathcal{Y}}^2+6 M_{\rm P}^4 H ^2 \Omega_{c}  \phi  ^3 \dot{\phi}  G_{3,\mathcal{YY}}+12 M_{\rm P}^4 H ^2 \Omega_{c}  \phi   \dot{\phi} G_{3,\mathcal{Y}} \,,
\end{eqnarray}

\begin{eqnarray}
    \Xi_{1} & \equiv & -2 a ^2G_{3,\mathcal{Y}}^4\dot{\phi}  f_{,\mathcal{E}} f_{,\mathcal{YE}} \phi  ^{11}-4 a ^3 H G_{3,\mathcal{Y}}^4 f_{,\mathcal{E}}^2 \phi^{10}-2 a ^2G_{3,\mathcal{Y}}^4\dot{\phi}  f_{,\mathcal{E}}^2 \phi  ^9+4 M_{\rm P}^2 a ^2G_{3,\mathcal{Y}}^4\dot{\phi}  f_{,\mathcal{E}}f_{,\mathcal{YE}} \phi  ^9 \nonumber \\
    & & +8 M_{\rm P}^2 a G_{3,\mathcal{Y}}^2\dot{\phi} ^2 G_{3,\mathcal{YY}} f_{,\mathcal{E}} f_{,\mathcal{YE}} \phi  ^9+3 M_{\rm P}^2 a ^3 H ^3 \Omega_{c} G_{3,\mathcal{Y}}^4 \phi  ^8-4 M_{\rm P}^2 a ^2G_{3,\mathcal{Y}}^2\dot{\phi}  G_{3,\mathcal{YY}} f_{,\mathcal{E}}^3 \phi  ^8+8 M_{\rm P}^2 a ^3 H G_{3,\mathcal{Y}}^4 f_{,\mathcal{E}}^2 \phi  ^8 \nonumber \\
    & & +16 M_{\rm P}^2 a ^2 H G_{3,\mathcal{Y}}^2\dot{\phi}  G_{3,\mathcal{YY}} f_{,\mathcal{E}}^2 \phi  ^8-4 M_{\rm P}^2 a ^2G_{3,\mathcal{Y}}^3\dot{\phi}  f_{,\mathcal{E}}^2f_{,\mathcal{YE}} \phi  ^8+8 M_{\rm P}^2 a ^2 H G_{3,\mathcal{Y}}^3\dot{\phi}  f_{,\mathcal{E}} f_{,\mathcal{YE}} \phi  ^8+4 M_{\rm P}^2 a ^2G_{3,\mathcal{Y}}\dot{\phi} G_{3,\mathcal{YY}} f_{,\mathcal{E}}^4 \phi  ^7 \nonumber \\
    & & -16 M_{\rm P}^2 a^3 H G_{3,\mathcal{Y}}^3 f_{,\mathcal{E}}^3 \phi^7+16 M_{\rm P}^2 a^3 H^2 G_{3,\mathcal{Y}}^3 f_{,\mathcal{E}}^2 \phi^7+8 M_{\rm P}^4 a G_{3,\mathcal{Y}} \dot{\phi}^2 G_{3,\mathcal{YY}}^2 f_{,\mathcal{E}}^2 \phi^7+8 M_{\rm P}^2 a^2G_{3,\mathcal{Y}}^4\dot{\phi}f_{,\mathcal{E}}^2 \phi^7 \nonumber \\
    & & +8 M_{\rm P}^2 aG_{3,\mathcal{Y}}^2\dot{\phi} ^2 G_{3,\mathcal{YY}} f_{,\mathcal{E}}^2 \phi  ^7-4 M_{\rm P}^4 a G_{3,\mathcal{Y}}^2\dot{\phi} ^2 G_{3,\mathcal{YYY}} f_{,\mathcal{E}}^2 \phi  ^7+16 M_{\rm P}^2 a G_{3,\mathcal{Y}}^3\dot{\phi} ^2 f_{,\mathcal{E}}f_{,\mathcal{YE}} \phi  ^7-8 M_{\rm P}^4\dot{\phi} ^3G_{3,\mathcal{YY}}^2 f_{,\mathcal{E}} f_{,\mathcal{YE}} \phi  ^7 \nonumber \\
    & & -8 M_{\rm P}^4 a G_{3,\mathcal{Y}}^2\dot{\phi} ^2 G_{3,\mathcal{YY}} f_{,\mathcal{E}}f_{,\mathcal{YE}} \phi  ^7-12 M_{\rm P}^4 a ^3 H ^3 \Omega_{c} G_{3,\mathcal{Y}}^4 \phi  ^6+8 M_{\rm P}^2 a ^3 H G_{3,\mathcal{Y}}^2 f_{,\mathcal{E}}^4 \phi  ^6-8 M_{\rm P}^4 a \dot{\phi} ^2 G_{3,\mathcal{YY}}^2 f_{,\mathcal{E}}^3 \phi  ^6 \nonumber \\
    & & -20 M_{\rm P}^2 a ^2G_{3,\mathcal{Y}}^3\dot{\phi} f_{,\mathcal{E}}^3 \phi  ^6+8 M_{\rm P}^4 a G_{3,\mathcal{Y}}\dot{\phi} ^2 G_{3,\mathcal{YYY}} f_{,\mathcal{E}}^3 \phi  ^6-16 M_{\rm P}^4 a  H  \dot{\phi}^2 G_{3,\mathcal{YY}}^2 f_{,\mathcal{E}}^2 \phi  ^6+40 M_{\rm P}^2 a ^2 H G_{3,\mathcal{Y}}^3 \dot{\phi} f_{,\mathcal{E}}^2 \phi  ^6 \nonumber \\
    & & -8 M_{\rm P}^4 a ^2 HG_{3,\mathcal{Y}}^2\dot{\phi} G_{3,\mathcal{YY}} f_{,\mathcal{E}}^2 \phi  ^6-4 M_{\rm P}^4 a G_{3,\mathcal{Y}}^2 G_{3,\mathcal{YY}} \ddot{\phi} f_{,\mathcal{E}}^2 \phi  ^6-12 M_{\rm P}^4 a ^2 H ^3 \Omega_{c}  G_{3,\mathcal{Y}}^2\dot{\phi}  G_{3,\mathcal{YY}} \phi  ^6 \nonumber \\
    & & +8 M_{\rm P}^4 a G_{3,\mathcal{Y}} \dot{\phi} ^2 G_{3,\mathcal{YY}} f_{,\mathcal{E}}^2 f_{,\mathcal{YE}} \phi  ^6-8 M_{\rm P}^4 a ^2 H G_{3,\mathcal{Y}}^3\dot{\phi}  f_{,\mathcal{E}}f_{,\mathcal{YE}} \phi  ^6-16 M_{\rm P}^4 a  H G_{3,\mathcal{Y}}\dot{\phi}^2 G_{3,\mathcal{YY}} f_{,\mathcal{E}} f_{,\mathcal{YE}} \phi  ^6 \nonumber \\
    & & +12 M_{\rm P}^2 a ^2G_{3,\mathcal{Y}}^2\dot{\phi}  f_{,\mathcal{E}}^4 \phi  ^5-4 M_{\rm P}^4 a \dot{\phi} ^2 G_{3,\mathcal{YYY}} f_{,\mathcal{E}}^4 \phi  ^5-12 M_{\rm P}^4 a ^3 H ^4 \Omega_{c} G_{3,\mathcal{Y}}^3 \phi  ^5+24 M_{\rm P}^4 a ^2 H G_{3,\mathcal{Y}}\dot{\phi}  G_{3,\mathcal{YY}} f_{,\mathcal{E}}^3 \phi  ^5 \nonumber \\
    & & +8 M_{\rm P}^4 a G_{3,\mathcal{Y}} G_{3,\mathcal{YY}} \dot{\phi}'  f_{,\mathcal{E}}^3 \phi  ^5-16 M_{\rm P}^4 a ^3 H ^2G_{3,\mathcal{Y}}^3 f_{,\mathcal{E}}^2 \phi  ^5+16 M_{\rm P}^2 a G_{3,\mathcal{Y}}^3\dot{\phi} ^2  f_{,\mathcal{E}}^2 \phi  ^5-8 M_{\rm P}^4\dot{\phi} ^3 G_{3,\mathcal{YY}}^2 f_{,\mathcal{E}}^2 \phi  ^5 \nonumber \\
    & & -4 M_{\rm P}^4 a ^2G_{3,\mathcal{Y}}^3 \dot{H} f_{,\mathcal{E}}^2 \phi  ^5-32 M_{\rm P}^4 a ^2 H ^2G_{3,\mathcal{Y}} \dot{\phi} G_{3,\mathcal{YY}} f_{,\mathcal{E}}^2 \phi  ^5+24 M_{\rm P}^4 a ^3 H ^3 \Omega_{c} G_{3,\mathcal{Y}}^3 f_{,\mathcal{E}} \phi  ^5+8 M_{\rm P}^4 a ^2 H G_{3,\mathcal{Y}}^2 \dot{\phi} f_{,\mathcal{E}}^2 f_{,\mathcal{YE}} \phi^5 \nonumber \\
    & & -16 M_{\rm P}^4 a G_{3,\mathcal{Y}}^3\dot{\phi} ^2 f_{,\mathcal{E}}f_{,\mathcal{YE}} \phi  ^5-8 M_{\rm P}^4 a ^2 H ^2 G_{3,\mathcal{Y}}^2\dot{\phi}  f_{,\mathcal{E}}f_{,\mathcal{YE}} \phi  ^5-32 M_{\rm P}^4G_{3,\mathcal{Y}}\dot{\phi} ^3 G_{3,\mathcal{YY}} f_{,\mathcal{E}}f_{,\mathcal{YE}} \phi  ^5+12 M_{\rm P}^6 a ^3 H ^3 \Omega_{c} G_{3,\mathcal{Y}}^4 \phi  ^4 \nonumber \\
    & & -16 M_{\rm P}^4 a ^2 H \dot{\phi} G_{3,\mathcal{YY}} f_{,\mathcal{E}}^4 \phi  ^4-4 M_{\rm P}^4 a  G_{3,\mathcal{YY}}\ddot{\phi}  f_{,\mathcal{E}}^4 \phi  ^4+32 M_{\rm P}^4 a ^3 H ^2G_{3,\mathcal{Y}}^2 f_{,\mathcal{E}}^3 \phi  ^4+8 M_{\rm P}^4 a ^2G_{3,\mathcal{Y}}^2 \dot{H}  f_{,\mathcal{E}}^3 \phi  ^4 \nonumber \\
    & & +24 M_{\rm P}^4 a G_{3,\mathcal{Y}}\dot{\phi} ^2 G_{3,\mathcal{YY}} f_{,\mathcal{E}}^3 \phi  ^4+12 M_{\rm P}^6 a  H ^3 \Omega_{c} \dot{\phi} ^2 G_{3,\mathcal{YY}}^2 \phi  ^4-16 M_{\rm P}^4 a ^3 H ^3G_{3,\mathcal{Y}}^2 f_{,\mathcal{E}}^2 \phi  ^4-12 M_{\rm P}^4 a ^3 H ^3 \Omega_{c} G_{3,\mathcal{Y}}^2 f_{,\mathcal{E}}^2 \phi  ^4 \nonumber \\
    & & -28 M_{\rm P}^4 a ^2 H G_{3,\mathcal{Y}}^3\dot{\phi}  f_{,\mathcal{E}}^2 \phi  ^4-80 M_{\rm P}^4 a  H G_{3,\mathcal{Y}}\dot{\phi} ^2 G_{3,\mathcal{YY}} f_{,\mathcal{E}}^2 \phi  ^4-8 M_{\rm P}^4 a G_{3,\mathcal{Y}}^3 \ddot{\phi}  f_{,\mathcal{E}}^2 \phi  ^4-24 M_{\rm P}^4 a ^2 H ^3 \Omega_{c}  G_{3,\mathcal{Y}}^3\dot{\phi}  \phi  ^4 \nonumber \\
    & & +24 M_{\rm P}^6 a ^2 H ^3 \Omega_{c} G_{3,\mathcal{Y}}^2\dot{\phi} G_{3,\mathcal{YY}} \phi  ^4+16 M_{\rm P}^4 a G_{3,\mathcal{Y}}^2\dot{\phi} ^2 f_{,\mathcal{E}}^2f_{,\mathcal{YE}} \phi  ^4-32 M_{\rm P}^4 a  HG_{3,\mathcal{Y}}^2\dot{\phi} ^2 f_{,\mathcal{E}} f_{,\mathcal{YE}} \phi  ^4-16 M_{\rm P}^4 a ^3 H ^2G_{3,\mathcal{Y}} f_{,\mathcal{E}}^4 \phi  ^3 \nonumber \\
    & & -4 M_{\rm P}^4 a ^2G_{3,\mathcal{Y}} \dot{H} f_{,\mathcal{E}}^4 \phi  ^3-24 M_{\rm P}^4 a \dot{\phi} ^2 G_{3,\mathcal{YY}} f_{,\mathcal{E}}^4 \phi  ^3+24 M_{\rm P}^6 a ^3 H ^4 \Omega_{c} G_{3,\mathcal{Y}}^3 \phi  ^3+64 M_{\rm P}^4 a ^2 H G_{3,\mathcal{Y}}^2\dot{\phi} f_{,\mathcal{E}}^3 \phi  ^3+16 M_{\rm P}^4 aG_{3,\mathcal{Y}}^2\ddot{\phi} f_{,\mathcal{E}}^3 \phi  ^3 \nonumber \\
    & & -72 M_{\rm P}^4 a ^2 H ^2G_{3,\mathcal{Y}}^2\dot{\phi} f_{,\mathcal{E}}^2 \phi  ^3-32 M_{\rm P}^4G_{3,\mathcal{Y}}\dot{\phi} ^3 G_{3,\mathcal{YY}} f_{,\mathcal{E}}^2 \phi  ^3+24 M_{\rm P}^6 a ^2 H ^4 \Omega_{c} G_{3,\mathcal{Y}}\dot{\phi}  G_{3,\mathcal{Y}} \phi  ^3-48 M_{\rm P}^6 a ^3 H ^3 \Omega_{c} G_{3,\mathcal{Y}}^3 f_{,\mathcal{E}} \phi  ^3 \nonumber \\
    & & -48 M_{\rm P}^6 a ^2 H ^3 \Omega_{c} G_{3,\mathcal{Y}}\dot{\phi} G_{3,\mathcal{YY}} f_{,\mathcal{E}} \phi  ^3-32 M_{\rm P}^4G_{3,\mathcal{Y}}^2\dot{\phi} ^3 f_{,\mathcal{E}}f_{,\mathcal{YE}} \phi ^3-36 M_{\rm P}^4 a ^2 H G_{3,\mathcal{Y}}\dot{\phi} f_{,\mathcal{E}}^4 \phi ^2-8 M_{\rm P}^4 a G_{3,\mathcal{Y}}\ddot{\phi}  f_{,\mathcal{E}}^4 \phi  ^2 \nonumber \\
    & & +16 M_{\rm P}^4 a G_{3,\mathcal{Y}}^2\dot{\phi} ^2 f_{,\mathcal{E}}^3 \phi  ^2+12 M_{\rm P}^6 a ^3 H ^5 \Omega_{c} G_{3,\mathcal{Y}}^2 \phi  ^2+72 M_{\rm P}^6 a ^3 H ^3 \Omega_{c}G_{3,\mathcal{Y}}^2 f_{,\mathcal{E}}^2 \phi  ^2-96 M_{\rm P}^4 a  HG_{3,\mathcal{Y}}^2\dot{\phi} ^2 f_{,\mathcal{E}}^2 \phi  ^2 \nonumber \\
    & & +24 M_{\rm P}^6 a ^2 H ^3 \Omega_{c} \dot{\phi}  G_{3,\mathcal{YY}} f_{,\mathcal{E}}^2 \phi  ^2+48 M_{\rm P}^6 a ^2 H ^3 \Omega_{c} G_{3,\mathcal{Y}}^3\dot{\phi}  \phi  ^2+48 M_{\rm P}^6 a  H ^3 \Omega_{c} G_{3,\mathcal{Y}}\dot{\phi} ^2 G_{3,\mathcal{YY}} \phi  ^2-48 M_{\rm P}^6 a ^3 H ^4 \Omega_{c} G_{3,\mathcal{Y}}^2 f_{,\mathcal{E}} \phi  ^2 \nonumber \\
    & & -16 M_{\rm P}^4 a G_{3,\mathcal{Y}}\dot{\phi} ^2 f_{,\mathcal{E}}^4 \phi  -48 M_{\rm P}^6 a ^3 H ^3 \Omega_{c} G_{3,\mathcal{Y}} f_{,\mathcal{E}}^3 \phi  -32 M_{\rm P}^4G_{3,\mathcal{Y}}^2\dot{\phi} ^3 f_{,\mathcal{E}}^2 \phi  +24 M_{\rm P}^6 a ^3 H ^4 \Omega_{c} G_{3,\mathcal{Y}} f_{,\mathcal{E}}^2 \phi  +48 M_{\rm P}^6  a ^2 H ^4 \Omega_{c} G_{3,\mathcal{Y}}^2\dot{\phi}  \phi \nonumber \\
    & & -96 M_{\rm P}^6 a ^2 H ^3 \Omega_{c}G_{3,\mathcal{Y}}^2\dot{\phi}  f_{,\mathcal{E}} \phi  +12 M_{\rm P}^6 a ^3 H ^3 \Omega_{c}  f_{,\mathcal{E}}^4+48 M_{\rm P}^6 a  H ^3 \Omega_{c}  G_{3,\mathcal{Y}}^2\dot{\phi} ^2+48 M_{\rm P}^6 a ^2 H ^3 \Omega_{c} G_{3,\mathcal{Y}}\dot{\phi} 
   f_{,\mathcal{E}}^2 \,,
\end{eqnarray}

\begin{eqnarray}
    \Xi_{2} & \equiv & -a ^3 G_{3,\mathcal{Y}}^4 f_{,\mathcal{E}}^2 \phi  ^{10}+3 M_{\rm P}^2 a ^3 H ^2 \Omega_{c}  G_{3,\mathcal{Y}}^4 \phi  ^8+2 M_{\rm P}^2 a ^3 G_{3,\mathcal{Y}}^4 f_{,\mathcal{E}}^2 \phi  ^8+4 M_{\rm P}^2 a ^2 G_{3,\mathcal{Y}}^2 \dot{\phi} G_{3,\mathcal{YY}} f_{,\mathcal{E}}^2 \phi  ^8-4 M_{\rm P}^2 a ^3 G_{3,\mathcal{Y}}^3 f_{,\mathcal{E}}^3 \phi  ^7 \nonumber \\
    & & +4 M_{\rm P}^2 a ^3 H G_{3,\mathcal{Y}}^3 f_{,\mathcal{E}}^2 \phi  ^7-12 M_{\rm P}^4 a ^3 H ^2 \Omega_{c}  G_{3,\mathcal{Y}}^4 \phi  ^6+2 M_{\rm P}^2 a ^3 G_{3,\mathcal{Y}}^2 f_{,\mathcal{E}}^4 \phi  ^6-4 M_{\rm P}^4 a  \dot{\phi} ^2 G_{3,\mathcal{YY}}^2 f_{,\mathcal{E}}^2 \phi  ^6+8 M_{\rm P}^2 a ^2 G_{3,\mathcal{Y}}^3 \dot{\phi} f_{,\mathcal{E}}^2 \phi  ^6 \nonumber \\
    & & -4 M_{\rm P}^4 a ^2 G_{3,\mathcal{Y}}^2 \dot{\phi} G_{3,\mathcal{YY}} f_{,\mathcal{E}}^2 \phi  ^6-12 M_{\rm P}^4 a ^2 H ^2 \Omega_{c}  G_{3,\mathcal{Y}}^2 \dot{\phi} G_{3,\mathcal{YY}} \phi  ^6-12 M_{\rm P}^4 a ^3 H ^3 \Omega_{c}  G_{3,\mathcal{Y}}^3 \phi  ^5+8 M_{\rm P}^4 a ^2 G_{3,\mathcal{Y}} \dot{\phi}  G_{3,\mathcal{YY}} f_{,\mathcal{E}}^3 \phi  ^5 \nonumber \\
    & & -4 M_{\rm P}^4 a ^3 H G_{3,\mathcal{Y}}^3 f_{,\mathcal{E}}^2 \phi  ^5-8 M_{\rm P}^4 a ^2 H G_{3,\mathcal{Y}} \dot{\phi}  G_{3,\mathcal{YY}} f_{,\mathcal{E}}^2 \phi  ^5+24 M_{\rm P}^4 a ^3 H ^2 \Omega_{c}  G_{3,\mathcal{Y}}^3 f_{,\mathcal{E}} \phi  ^5+12 M_{\rm P}^6 a ^3 H ^2 \Omega_{c}  G_{3,\mathcal{Y}}^4 \phi  ^4 \nonumber \\
    & & -4 M_{\rm P}^4 a ^2 \dot{\phi}  G_{3,\mathcal{YY}} f_{,\mathcal{E}}^4 \phi  ^4+8 M_{\rm P}^4 a ^3 H  G_{3,\mathcal{Y}}^2 f_{,\mathcal{E}}^3 \phi  ^4+12 M_{\rm P}^6 a  H ^2 \Omega_{c}  \dot{\phi} ^2 G_{3,\mathcal{YY}}^2 \phi  ^4-4 M_{\rm P}^4 a ^3 H ^2 G_{3,\mathcal{Y}}^2 f_{,\mathcal{E}}^2 \phi  ^4 \nonumber \\
    & & -12 M_{\rm P}^4 a ^3 H ^2 \Omega_{c} G_{3,\mathcal{Y}}^2 f_{,\mathcal{E}}^2 \phi  ^4-8 M_{\rm P}^4 a ^2 G_{3,\mathcal{Y}}^3 \dot{\phi}  f_{,\mathcal{E}}^2 \phi  ^4-16 M_{\rm P}^4 a G_{3,\mathcal{Y}} \dot{\phi} ^2 G_{3,\mathcal{YY}} f_{,\mathcal{E}}^2 \phi  ^4-24 M_{\rm P}^4 a ^2 H ^2 \Omega_{c}  G_{3,\mathcal{Y}}^3 \dot{\phi}  \phi  ^4 \nonumber \\
    & & +24 M_{\rm P}^6 a ^2 H ^2 \Omega_{c}  G_{3,\mathcal{Y}}^2 \dot{\phi}  G_{3,\mathcal{YY}} \phi^4-4 M_{\rm P}^4 a ^3 H  G_{3,\mathcal{Y}} f_{,\mathcal{E}}^4 \phi  ^3+24 M_{\rm P}^6 a ^3 H ^3 \Omega_{c}  G_{3,\mathcal{Y}}^3 \phi  ^3 +16 M_{\rm P}^4 a ^2 G_{3,\mathcal{Y}}^2 \dot{\phi}  f_{,\mathcal{E}}^3 \phi  ^3 \nonumber \\
    & & -16 M_{\rm P}^4 a ^2 H  G_{3,\mathcal{Y}}^2 \dot{\phi}  f_{,\mathcal{E}}^2 \phi  ^3+24 M_{\rm P}^6 a ^2 H ^3 \Omega_{c} G_{3,\mathcal{Y}} \dot{\phi}  G_{3,\mathcal{YY}} \phi  ^3-48 M_{\rm P}^6 a ^3 H ^2 \Omega_{c}  G_{3,\mathcal{Y}}^3 f_{,\mathcal{E}} \phi  ^3-48 M_{\rm P}^6 a ^2H ^2 \Omega_{c}  G_{3,\mathcal{Y}} \dot{\phi}  G_{3,\mathcal{YY}} f_{,\mathcal{E}} \phi  ^3 \nonumber \\
    & & -8 M_{\rm P}^4 a ^2 G_{3,\mathcal{Y}} \dot{\phi}  f_{,\mathcal{E}}^4 \phi  ^2+12 M_{\rm P}^6 a ^3 H ^4 \Omega_{c}  G_{3,\mathcal{Y}}^2 \phi  ^2+72 M_{\rm P}^6 a ^3 H ^2 \Omega_{c}  G_{3,\mathcal{Y}}^2 f_{,\mathcal{E}}^2 \phi  ^2-16 M_{\rm P}^4 a  G_{3,\mathcal{Y}}^2 \dot{\phi} ^2 f_{,\mathcal{E}}^2 \phi  ^2 \nonumber \\
    & & +24 M_{\rm P}^6 a ^2 H ^2 \Omega_{c}  \dot{\phi}  G_{3,\mathcal{YY}} f_{,\mathcal{E}}^2 \phi^2+48 M_{\rm P}^6 a ^2 H ^2 \Omega_{c}  G_{3,\mathcal{Y}}^3 \dot{\phi}  \phi  ^2+48 M_{\rm P}^6 a  H ^2 \Omega_{c}  G_{3,\mathcal{Y}} \dot{\phi} ^2 G_{3,\mathcal{YY}} \phi  ^2-48 M_{\rm P}^6 a ^3 H ^3 \Omega_{c}  G_{3,\mathcal{Y}}^2 f_{,\mathcal{E}} \phi  ^2 \nonumber \\
    & & -48 M_{\rm P}^6 a ^3 H ^2 \Omega_{c}  G_{3,\mathcal{Y}} f_{,\mathcal{E}}^3 \phi  +24 M_{\rm P}^6 a ^3 H ^3 \Omega_{c}  G_{3,\mathcal{Y}} f_{,\mathcal{E}}^2 \phi +48 M_{\rm P}^6 a ^2 H ^3 \Omega_{c}  G_{3,\mathcal{Y}}^2 \dot{\phi}  \phi  -96 M_{\rm P}^6 a ^2 H ^2  \Omega_{c}  G_{3,\mathcal{Y}}^2 \dot{\phi}f_{,\mathcal{E}} \phi \nonumber \\
    & & +12 M_{\rm P}^6 a ^3 H ^2 \Omega_{c}  f_{,\mathcal{E}}^4+48 M_{\rm P}^6 a  H ^2 \Omega_{c} G_{3,\mathcal{Y}}^2 \dot{\phi} ^2+48 M_{\rm P}^6 a ^2 H ^2 \Omega_{c}  G_{3,\mathcal{Y}} \dot{\phi}  f_{,\mathcal{E}}^2 \,.
\end{eqnarray}

\bibliographystyle{JHEP}
\bibliography{bibliography}

\end{document}